%% file: artitd47_rapport_v4_no_comment.tex
\documentclass[twoside,11pt,a4paper]{article}
\begin{document}

\input{bss_com}

\input{artitd47_rapport_v4_com}
\normalsize
\sloppy
\begin{center}
{
\Large
\bf
Effect of indirect dependencies on
"Maximum likelihood blind separation of two quantum states (qubits)
with cylindrical-%
symmetry
Heisenberg spin coupling"
}
~\\
~\\
Yannick Deville$^1$ 
and
Alain Deville$^2$ 
~\\
~\\
(1) 
Laboratoire
d'Astrophysique de Toulouse-Tarbes,
Universit\'e de Toulouse, CNRS,
14 Av. Edouard Belin, 31400 Toulouse, France.
Email:
ydeville@ast.obs-mip.fr
~\\
(2) 
IM2NP, Universit\'e de Provence, Centre de
Saint-J\'er\^ome, 13397 Marseille Cedex 20, France.
Email: alain.deville@univ-provence.fr
\end{center}
~\\
~\\
~\\
{\bf Abstract.}
In a previous paper
\cite{amoi5-43},
we investigated the Blind Source Separation (BSS) problem, for
the nonlinear mixing model that we introduced 
in that paper.
We proposed to solve this problem by using a maximum likelihood
(ML) approach.
When 
applying
the ML approach to BSS problems,
one usually 
determines
the analytical expressions of
the 
derivatives
of the log-likelihood with respect to the parameters of
the considered mixing model.
In the literature,
these calculations
were mainly considered 
for 
linear mixtures
up to now. They are 
more complex for nonlinear mixtures,
due to 
dependencies 
between the considered quantities.
Moreover, the notations commonly employed by the BSS community in such
calculations may become misleading when using them for
nonlinear mixtures, due to the above-mentioned dependencies.
In this document, we therefore explain this phenomenon, by showing
the effect of indirect dependencies
on the application of the ML approach to the mixing model
considered in
\cite{amoi5-43}.
This yields the explicit expression of the 
complete derivative of the
log-likelihood associated to that mixing model.
~\\
~\\
{\bf Keywords.}
Maximum likelihood estimation,
blind signal separation, 
quantum source separation,
nonlinear mixtures,
indirect dependency,
total derivative,
partial derivative,
gradient.
~\\
~\\
\section{Data model}
In a
previous paper
\cite{amoi5-43}, we investigated a Blind Quantum Source (or Signal)
Separation (BQSS) problem.
More precisely, we aimed at restoring two quantum states (qubits) after
they have been coupled, i.e. after they have
been "mixed", using the classical
Blind Source 
Separation (BSS) terminology.
The considered coupling was based on a cylindrical-symmetry Heisenberg
model.

We showed that repeated initializations
(i.e. preparations)
and measurements performed with these
coupled qubits
resulted in an "observation vector" 
which may be denoted
$
\Yssmixsigdisctimecentvec
=
[
\Yssmixsigdisctimecentone
,
\Yssmixsigdisctimecenttwo
,
\Yssmixsigdisctimecentthree
 ] ^{T}
$
using standard BSS notations, where
$^{T}$
stands for transpose.
In the considered problem, the 
components
of this vector are equal to
$
\Yssmixsigdisctimecentone = \Ytwoqubitsprobaplusplus,
\Yssmixsigdisctimecenttwo = \Ytwoqubitsprobaminusminus
$
and $ \Yssmixsigdisctimecentthree = \Ytwoqubitsprobaplusminus$,
where the quantities \ytwoqubitsprobaindexstd\ are defined in
\cite{amoi5-43}.

Moreover, we proved
that the 
components
of the above
observation vector may be expressed as nonlinear combinations
(i.e. nonlinear "mixtures", in BSS terms) 
of a set of
"source signals".
Using standard BSS notations, the vector composed of these source
signals reads
$
\Ysssrcsigdisctimecentvec
=
[
\Ysssrcsigdisctimecentone
,
\Ysssrcsigdisctimecenttwo
,
\Ysssrcsigdisctimecentthree
 ] ^{T}
$.
In the considered problem, the 
components
of this vector are equal to
$
\Ysssrcsigdisctimecentone = \Yparamqubitonestateplusmodulus,
\Ysssrcsigdisctimecenttwo = \Yparamqubittwostateplusmodulus
$
and $ \Ysssrcsigdisctimecentthree = \Ytwoqubitresultphaseinit $,
where the right-hand terms of these equations are defined in
\cite{amoi5-43}.

The 
"mixing model"
then consists of the equations which define how the 
components
of
the observation vector are expressed with respect to (i) the 
components
of
the source vector and (ii) the mixing parameter(s).
By modelling the considered quantum 
configuration, we showed in
\cite{amoi5-43}
that this configuration involves a single mixing parameter,
denoted
\ytwoqubitresultphaseevolsin .
The mixing equations then read
(see (18), (24), (25) in \cite{amoi5-43})
\begin{eqnarray}
\label{eq-statecoef-vs-twoqubitsprobaplusplus-polar}
\Yparamqubitonestateplusmodulus ^2 \Yparamqubittwostateplusmodulus
^2 &
=
& \Ytwoqubitsprobaplusplus
\\
\label{eq-statecoef-vs-twoqubitsprobaminusminus-polar-vs-paramqubitindexstdstateplusmodulus}
( 1
-
\Yparamqubitonestateplusmodulus ^2
)
(
1
-
\Yparamqubittwostateplusmodulus ^2
)
&
=
&
\Ytwoqubitsprobaminusminus
\\
\nonumber
\Yparamqubitonestateplusmodulus ^2
(
1
-
\Yparamqubittwostateplusmodulus ^2
)
(
1 -
\Ytwoqubitresultphaseevolsin ^2
)
+
(
1
-
\Yparamqubitonestateplusmodulus ^2
)
\Yparamqubittwostateplusmodulus ^2
\Ytwoqubitresultphaseevolsin ^2
\hspace{30mm}
\\
\label{eq-statecoef-vs-twoqubitsprobaplusminus-polar-versionthree}
-
2 \Yparamqubitonestateplusmodulus \Yparamqubittwostateplusmodulus
\sqrt{
1
-
\Yparamqubitonestateplusmodulus ^2
}
\sqrt{
1
-
\Yparamqubittwostateplusmodulus ^2
}
\sqrt{
1 -
\Ytwoqubitresultphaseevolsin ^2
}
\Ytwoqubitresultphaseevolsin
\sin
\Ytwoqubitresultphaseinit
&
=
&
\Ytwoqubitsprobaplusminus .
\end{eqnarray}

This mixing model may also be expressed in
compact form as
\begin{equation}
\label{eq-mix-model-vector}
\Yssmixsigdisctimecentvec
=
\Ymixfunc
(
\Ysssrcsigdisctimecentvec
)
\end{equation}
where 
the
nonlinear mixing function
\ymixfunc\
has
three 
components
$
\Ymixfunc _{1}
$
to
$
\Ymixfunc _{3}
$,
with
$
\Yssmixsigdisctimecentindex
=
\Ymixfunc _{\Ysswayindex}
(
\Ysssrcsigdisctimecentvec
),
$
$
\
\forall 
\Ysswayindex \in \{ 1 \dots 3 \}
$.
These 
components 
$
\Ymixfunc _{\Ysswayindex}
$
are
respectively defined by
(\ref{eq-statecoef-vs-twoqubitsprobaplusplus-polar}),
(\ref{eq-statecoef-vs-twoqubitsprobaminusminus-polar-vs-paramqubitindexstdstateplusmodulus})
and
(\ref{eq-statecoef-vs-twoqubitsprobaplusminus-polar-versionthree}).

Eq.
(\ref{eq-mix-model-vector}) focuses on the signals (i.e. sources
and observations).
It hides the fact that the observations also depend on the
parameters of the mixing model, i.e. on 
\ytwoqubitresultphaseevolsin\
in the model considered here.
This additional dependency can be made explicit, by rewriting
(\ref{eq-mix-model-vector}) as
\begin{equation}
\label{eq-mix-model-vector-with-mix-param}
\Yssmixsigdisctimecentvec
=
\Ymixfunc
(
\Ysssrcsigdisctimecentvec ,
\Ytwoqubitresultphaseevolsin
)
.
\end{equation}
The latter form of the mixing model is better suited to the 
maximum likelihood approach considered below in this document.
\section{Previously reported results for maximum likelihood approach}
\label{sec-ml-previous}

In 
our BSS problem,
we aim at retrieving a sequence of
unknown source vectors 
\ysssrcsigdisctimecentvec\
from the corresponding sequence
of measured observation vectors
\yssmixsigdisctimecentvec\ 
and from
the 
mixing parameter
\ytwoqubitresultphaseevolsin ,
which
is also initially unknown.
This parameter 
\ytwoqubitresultphaseevolsin\
should therefore be estimated before proceeding to the source
restoration step.

In 
\cite{amoi5-43},
we investigated the estimation of 
\ytwoqubitresultphaseevolsin\
by means of the maximum likelihood (ML)
approach.
While we detailed this procedure in 
\cite{amoi5-43},
we here only summarize its features which are of importance
for
developing new aspects of this ML approach further in this document.

The 
function
used to estimate 
\ytwoqubitresultphaseevolsin\
is the (normalized)
log-likelihood of the considered data.
Under some assumptions \cite{amoi5-43},
the log-likelihood here reads
(see (34) in \cite{amoi5-43})
\begin{equation}
\label{eq-ln-pdf-mix-vs-src-versiontwo}
\Ymlcostlnmean
=
\sum_{\Ysswayindex = 1}^{3}
E_t[
\ln
\Ysssrcsigdisctimecentindexranddens
( \Ysssrcsigdisctimecentindex (t) )
]
-
E_t[
\ln
|
\Ymixfuncjacob
( \Ysssrcsigdisctimecentvec (t) )
|
]
\end{equation}
where
$E_t[.]$
represents 
temporal averaging over the sequence
of available data,
$
\Ysssrcsigdisctimecentindexranddens
( . )
$
are the probability density functions of the source signals
and
\ymixfuncjacobval\
is the
Jacobian of the 
mixing function
\ymixfunc .
For the function
\ymixfunc\ considered in this investigation, 
we have (see (28) in \cite{amoi5-43})
\begin{equation}
\label{eq-mixfuncjacobval}
\Ymixfuncjacobval
=
8
\Yparamqubitonestateplusmodulus ^2
\Yparamqubittwostateplusmodulus ^2
(
\Yparamqubittwostateplusmodulus ^2
-
\Yparamqubitonestateplusmodulus ^2
)
\sqrt{
1
-
\Yparamqubitonestateplusmodulus ^2
}
\sqrt{
1
-
\Yparamqubittwostateplusmodulus ^2
}
\sqrt{
1 -
\Ytwoqubitresultphaseevolsin ^2
}
\Ytwoqubitresultphaseevolsin
\cos
\Ytwoqubitresultphaseinit
.
\end{equation}

When applying the ML approach to a
parameter estimation
problem,
the value selected for
the set of parameters to be estimated
is the value 
which maximizes
the log-likelihood
\ymlcostlnmean .
In order to determine 
the location
of this maximum, one usually considers the gradient of
\ymlcostlnmean\
( see e.g. \cite{icabook-oja}).
In our configuration, only a single parameter
is to be estimated, namely
\ytwoqubitresultphaseevolsin .
Therefore, the gradient
of
\ymlcostlnmean\
is here restricted to the derivative
of
\ymlcostlnmean\
with respect to 
\ytwoqubitresultphaseevolsin .
In 
\cite{amoi5-43}, we denoted this gradient or derivative
by using the notation
most often 
employed
in the BSS community (see e.g.
\cite{icabook-oja}),
i.e.
$
\frac{\partial \Ymlcostlnmean}{\partial \Ytwoqubitresultphaseevolsin}
$.
We keep this notation in this section, in order to clearly refer to
the equations available in
\cite{amoi5-43},
but in Section
\ref{sec-ml-new} we will show 
that it may be misleading and we will therefore introduce another
notation in Section
\ref{sec-ml-new}.
In
\cite{amoi5-43},
we showed 
that this gradient reads
(see (37) in \cite{amoi5-43})
\begin{equation}
\label{eq-costlnmean-gradient}
\frac{\partial \Ymlcostlnmean}{\partial \Ytwoqubitresultphaseevolsin}
=
-
\sum_{\Ysswayindex = 1}^{3}
E_t[
\psi_{\Ysswayindex} (\Ysssrcsigdisctimecentindex )
\frac{\partial
\Ysssrcsigdisctimecentindex
}{\partial \Ytwoqubitresultphaseevolsin}]
-E_t[
\frac{1}{\Ymixfuncjacob}
\frac{
\partial \Ymixfuncjacob
}{
\partial \Ytwoqubitresultphaseevolsin
}
]
\end{equation}
where
(see (36) in \cite{amoi5-43})
\begin{equation}
\psi_{\Ysswayindex}(u)
=
-\frac{\partial \ln{f_{\Ysssrcsigdisctimecentindexrand}(u)}}{\partial u}
\hspace{5mm}
\forall
\Ysswayindex
\in \{ 1 \dots 3 \}
\end{equation}
are the score functions of the source signals.

The last stage of this investigation consists
in deriving the expressions of all the terms of
the right-hand part of
(\ref{eq-costlnmean-gradient}).
In 
\cite{amoi5-43}, we showed 
that the terms
$
\displaystyle
E_t[
\psi_{\Ysswayindex} (\Ysssrcsigdisctimecentindex )
\frac{\partial
\Ysssrcsigdisctimecentindex
}{\partial \Ytwoqubitresultphaseevolsin}]
$
corresponding to
$
\Ysswayindex = 1
$
and
$
\Ysswayindex = 2
$
are equal to zero.
The term 
corresponding to
$
\Ysswayindex = 3
$
was derived from 
equations (38) to (40) in
\cite{amoi5-43},
which read
\begin{equation}
\label{eq-impldiff}
\frac{\partial \Ymixfuncimpldiff}
{\partial \Ysssrcsigdisctimecentthree}
\frac{\partial \Ysssrcsigdisctimecentthree}
{\partial \Ytwoqubitresultphaseevolsin}
+
\frac{\partial \Ymixfuncimpldiff}
{\partial \Ytwoqubitresultphaseevolsin}
=
0
\end{equation}
with
\begin{equation}
\frac{\partial \Ymixfuncimpldiff}
{\partial \Ysssrcsigdisctimecentthree}
=
- 2 \Yparamqubitonestateplusmodulus \Yparamqubittwostateplusmodulus
\sqrt{
1
-
\Yparamqubitonestateplusmodulus ^2
}
\sqrt{
1
-
\Yparamqubittwostateplusmodulus ^2
}
\sqrt{
1 -
\Ytwoqubitresultphaseevolsin ^2
}
\Ytwoqubitresultphaseevolsin
\cos
\Ytwoqubitresultphaseinit
\end{equation}
and
\begin{equation}
\frac{\partial \Ymixfuncimpldiff}
{\partial \Ytwoqubitresultphaseevolsin}
=
2
\Ytwoqubitresultphaseevolsin
(
\Yparamqubittwostateplusmodulus ^2
-
\Yparamqubitonestateplusmodulus ^2
)
-
2 \Yparamqubitonestateplusmodulus \Yparamqubittwostateplusmodulus
\sqrt{
1
-
\Yparamqubitonestateplusmodulus ^2
}
\sqrt{
1
-
\Yparamqubittwostateplusmodulus ^2
}
\frac
{
1 -
2 \Ytwoqubitresultphaseevolsin ^2
}
{
\sqrt{
1 -
\Ytwoqubitresultphaseevolsin ^2
}
}
\sin
\Ytwoqubitresultphaseinit
.
\end{equation}

For the sake of clarity, we here provide the resulting expression
of 
$
\frac{\partial \Ysssrcsigdisctimecentthree}
{\partial \Ytwoqubitresultphaseevolsin}
$,
which is also used below in the current document.
The above equations yield
\begin{eqnarray}
\frac{\partial \Ysssrcsigdisctimecentthree}
{\partial \Ytwoqubitresultphaseevolsin}
&
=
&
-
\left[
\frac{\partial \Ymixfuncimpldiff}
{\partial \Ysssrcsigdisctimecentthree}
\right]^{-1}
\frac{\partial \Ymixfuncimpldiff}
{\partial \Ytwoqubitresultphaseevolsin}
\\
\label{eq-deriv-sssrcsigdisctimecentthree-vs-twoqubitresultphaseevolsin}
&
=
&
\frac{
(
\Yparamqubittwostateplusmodulus ^2
-
\Yparamqubitonestateplusmodulus ^2
)
}
{
\Yparamqubitonestateplusmodulus
\Yparamqubittwostateplusmodulus
\sqrt{
1
-
\Yparamqubitonestateplusmodulus ^2
}
\sqrt{
1
-
\Yparamqubittwostateplusmodulus ^2
}
\sqrt{
1 -
\Ytwoqubitresultphaseevolsin ^2
}
\cos
\Ytwoqubitresultphaseinit
}
-
\frac
{
(1 -
2 \Ytwoqubitresultphaseevolsin ^2)
\sin
\Ytwoqubitresultphaseinit
}
{
(1 -
\Ytwoqubitresultphaseevolsin ^2
)
\Ytwoqubitresultphaseevolsin
\cos
\Ytwoqubitresultphaseinit
}
.
\end{eqnarray}

The last term that should be 
determined
to obtain the complete expression in
(\ref{eq-costlnmean-gradient}),
i.e. in (37) of \cite{amoi5-43},
is its term
$
\displaystyle
\frac{
\partial \Ymixfuncjacob
}{
\partial \Ytwoqubitresultphaseevolsin
}
$.
In Equation (41) of
\cite{amoi5-43},
we provided an explicit expression that we denoted
$
\displaystyle
\frac{
\partial \Ymixfuncjacob
}{
\partial \Ytwoqubitresultphaseevolsin
}
$.
We here aim at warning the reader that, although each of the equations
(37) and (41) of
\cite{amoi5-43} is correct in itself
if interpreted correctly, there may be a misunderstanding when
considering these equations together, because the notation
$
\displaystyle
\frac{
\partial \Ymixfuncjacob
}{
\partial \Ytwoqubitresultphaseevolsin
}
$
does not have the same meaning in both of them.
Briefly,
$
\displaystyle
\frac{
\partial \Ymixfuncjacob
}{
\partial \Ytwoqubitresultphaseevolsin
}
$
represents a total derivative in
(37) of
\cite{amoi5-43},
but a partial derivative
in (41) of
\cite{amoi5-43},
as detailed below in Section
\ref{sec-ml-new}.
The right-hand expression in
Equation (41) of
\cite{amoi5-43}
is therefore \emph{only one of the terms} which compose the
complete expression of
$
\displaystyle
\frac{
\partial \Ymixfuncjacob
}{
\partial \Ytwoqubitresultphaseevolsin
}
$
to be used in
(37) of
\cite{amoi5-43}.
In the following section of the current document, 
we clarify this point
and we determine
the 
complete expression of
$
\displaystyle
\frac{
\partial \Ymixfuncjacob
}{
\partial \Ytwoqubitresultphaseevolsin
}
$
of (37) of
\cite{amoi5-43},
i.e. of
(\ref{eq-costlnmean-gradient}) of the current document.
\section{New results for maximum likelihood approach}
\label{sec-ml-new}
When applying the ML approach to any BSS configuration,
the log-likelihood
\ymlcostlnmean\
is considered for the fixed set of observed vectors.
The only independent
variable in this approach is the set of mixing parameters
to be estimated,
which is here restricted to \ytwoqubitresultphaseevolsin .
The source vectors are
dependent variables, 
here linked to the observations and to
\ytwoqubitresultphaseevolsin\
by
(\ref{eq-mix-model-vector-with-mix-param}).
The overall variations of the log-likelihood
\ymlcostlnmean\
with respect to 
\ytwoqubitresultphaseevolsin\
result from 
two types of terms contained in the expression of
\ymlcostlnmean , i.e. (i) the terms involving 
\ytwoqubitresultphaseevolsin\ itself and
(ii) the terms involving 
the source signals
$
\Ysssrcsigdisctimecentone ,
\Ysssrcsigdisctimecenttwo
$
and $ \Ysssrcsigdisctimecentthree $,
which are here considered
as functions of
\ytwoqubitresultphaseevolsin\
and may therefore be denoted as
$
\Ysssrcsigdisctimecentone (\Ytwoqubitresultphaseevolsin),
\Ysssrcsigdisctimecenttwo (\Ytwoqubitresultphaseevolsin)
$
and
$
\Ysssrcsigdisctimecentthree (\Ytwoqubitresultphaseevolsin)
$
for the sake of clarity.

This approach should be kept in mind when interpreting all equations in
\cite{amoi5-43},
which were partly gathered in
Section
\ref{sec-ml-previous} of the current document.
Especially, the log-likelihood itself, which appears in the left-hand
term of
(\ref{eq-ln-pdf-mix-vs-src-versiontwo}),
may be denoted as
\ymlcostlnmeanargmixparam\
for the sake of clarity.
In order to determine the location
of the maximum of
this log-likelihood,
one should then 
consider
the \emph{total} derivative of
\ymlcostlnmeanargmixparam\
with respect to
\ytwoqubitresultphaseevolsin .
The notations with partial derivatives
in
(\ref{eq-costlnmean-gradient})
may therefore be misleading, as confirmed below.
Therefore,
(\ref{eq-costlnmean-gradient})
should preferably 
be rewritten as
\begin{equation}
\label{eq-costlnmean-gradient-versiontwo}
\frac{d \Ymlcostlnmean}
{d \Ytwoqubitresultphaseevolsin}
=
-
\sum_{\Ysswayindex = 1}^{3}
E_t[
\psi_{\Ysswayindex} (\Ysssrcsigdisctimecentindex )
\frac{d
\Ysssrcsigdisctimecentindex
}{d \Ytwoqubitresultphaseevolsin}]
-E_t[
\frac{1}{\Ymixfuncjacob}
\frac{
d \Ymixfuncjacob
}{
d \Ytwoqubitresultphaseevolsin
}
]
\end{equation}
with
\begin{equation}
\psi_{\Ysswayindex}(u)
=
-\frac{d \ln{f_{\Ysssrcsigdisctimecentindexrand}(u)}}{d u}
\hspace{5mm}
\forall
\Ysswayindex
\in \{ 1 \dots 3 \}
.
\end{equation}

The term
$
\displaystyle
\frac{
d \Ymixfuncjacob
}{
d \Ytwoqubitresultphaseevolsin
}
$
in
(\ref{eq-costlnmean-gradient-versiontwo})
then deserves some care because, as shown by
(\ref{eq-mixfuncjacobval}), the Jacobian
\ymixfuncjacob\ contains 
the above-defined
two types of dependencies
with respect to 
\ytwoqubitresultphaseevolsin ,
i.e.
(i) \emph{direct dependencies} due to the factors in
(\ref{eq-mixfuncjacobval}) which explicitly contain
\ytwoqubitresultphaseevolsin\
and
(ii) \emph{indirect 
dependencies} due to the factors in
(\ref{eq-mixfuncjacobval}) which depend on the source signals,
which themselves depend\footnote{%
When applying the ML approach to a general mixing model,
all source
signals depend on all mixing parameters .
For the specific mixing model considered here, we will show
below that \emph{only one} of the three source signals actually
depends on the mixing parameter.}
on
\ytwoqubitresultphaseevolsin\ in the ML approach.
We here have to consider the \emph{total}
derivative
$
\displaystyle
\frac{
d \Ymixfuncjacob
}{
d \Ytwoqubitresultphaseevolsin
}
$, which takes into account both types of dependencies, and which
therefore reads
\begin{equation}
\label{eq-cost-deriv-total}
\frac{
d \Ymixfuncjacob
}{
d \Ytwoqubitresultphaseevolsin
}
=
\frac{
\partial \Ymixfuncjacob
}{
\partial \Ytwoqubitresultphaseevolsin
}
+
\sum_{\Ysswayindex = 1}^{3}
\frac
{\partial \Ymixfuncjacob}
{\partial \Ysssrcsigdisctimecentindex}
\frac{d
\Ysssrcsigdisctimecentindex
}{d \Ytwoqubitresultphaseevolsin}
.
\end{equation}
In this expression,
$
\displaystyle
\frac{
\partial \Ymixfuncjacob
}{
\partial \Ytwoqubitresultphaseevolsin
}
$
is the \emph{partial} derivative of
\ymixfuncjacob\ with respect to
\ytwoqubitresultphaseevolsin ,
calculated
by considering that the source signals are constant.
This 
partial derivative is the quantity that
we 
provided
in Equation (41) of
\cite{amoi5-43},
where we also denoted it as
$
\displaystyle
\frac{
\partial \Ymixfuncjacob
}{
\partial \Ytwoqubitresultphaseevolsin
}
$.
However, 
since we independently used
the same
notation
$
\displaystyle
\frac{
\partial \Ymixfuncjacob
}{
\partial \Ytwoqubitresultphaseevolsin
}
$
in the last term of
(37) of
\cite{amoi5-43} 
(that we 
repeat
as
(\ref{eq-costlnmean-gradient})
of the current
document),
we may have incorrectly suggested to the reader that
this last term of
(37) of
\cite{amoi5-43}
is equal to the above-mentioned  partial derivative.
Instead, let us insist again that
the partial derivative
$
\displaystyle
\frac{
\partial \Ymixfuncjacob
}{
\partial \Ytwoqubitresultphaseevolsin
}
$
in
(41) of
\cite{amoi5-43}
is first to be added with the other terms in the right-hand
part of
(\ref{eq-cost-deriv-total}),
in order to obtain
the overall
total
derivative
$
\displaystyle
\frac{
d \Ymixfuncjacob
}{
d \Ytwoqubitresultphaseevolsin
}
$
defined by
(\ref{eq-cost-deriv-total}).
What should eventually 
be used in the last term of
(\ref{eq-costlnmean-gradient}) or
(\ref{eq-costlnmean-gradient-versiontwo})
is this \emph{total} derivative.

So, back to the 
calculation
of all terms of
the total derivative
$
\displaystyle
\frac{
d \Ymixfuncjacob
}{
d \Ytwoqubitresultphaseevolsin
}
$
in
(\ref{eq-cost-deriv-total}), the first term is available from
Equation (41) of
\cite{amoi5-43} and reads
\begin{equation}
\label{eq-mixfuncjacobval-gradient-partial}
\frac{\partial 
\Ymixfuncjacob
}
{\partial \Ytwoqubitresultphaseevolsin}
=
8
\Yparamqubitonestateplusmodulus ^2
\Yparamqubittwostateplusmodulus ^2
(
\Yparamqubittwostateplusmodulus ^2
-
\Yparamqubitonestateplusmodulus ^2
)
\sqrt{
1
-
\Yparamqubitonestateplusmodulus ^2
}
\sqrt{
1
-
\Yparamqubittwostateplusmodulus ^2
}
\frac
{
1 -
2 \Ytwoqubitresultphaseevolsin ^2
}
{
\sqrt{
1 -
\Ytwoqubitresultphaseevolsin ^2
}
}
\cos
\Ytwoqubitresultphaseinit
.
\end{equation}

The other three terms 
of
(\ref{eq-cost-deriv-total})
involve the derivatives\footnote{Unlike 
$
\frac{\partial 
\Ymixfuncjacob
}
{\partial \Ytwoqubitresultphaseevolsin}
$
considered above, the derivatives
$
\frac{d
\Ysssrcsigdisctimecentindex
}{d \Ytwoqubitresultphaseevolsin}
$
do not yield any risk of
ambiguity between partial and total
derivatives:
each considered signal
\ysssrcsigdisctimecentindex\ is here considered independently from
the other source signals and only involves a direct dependency with
respect to
\ytwoqubitresultphaseevolsin .}
$
\frac{d
\Ysssrcsigdisctimecentindex
}{d \Ytwoqubitresultphaseevolsin}
$.
Their calculation
first 
require us to determine
the expressions of the
source signals
$
\Ysssrcsigdisctimecentone = \Yparamqubitonestateplusmodulus,
\Ysssrcsigdisctimecenttwo = \Yparamqubittwostateplusmodulus
$
and $ \Ysssrcsigdisctimecentthree = \Ytwoqubitresultphaseinit $
with respect to the observations and
mixing parameter
\ytwoqubitresultphaseevolsin .
The first two source signals,
i.e.
$
\Ysssrcsigdisctimecentone = \Yparamqubitonestateplusmodulus$
and
$
\Ysssrcsigdisctimecenttwo = \Yparamqubittwostateplusmodulus
$
are obtained by solving
(\ref{eq-statecoef-vs-twoqubitsprobaplusplus-polar})
and
(\ref{eq-statecoef-vs-twoqubitsprobaminusminus-polar-vs-paramqubitindexstdstateplusmodulus}).
These equations are independent of
\ytwoqubitresultphaseevolsin .
Therefore,
$
\Ysssrcsigdisctimecentone
$
and
$
\Ysssrcsigdisctimecenttwo 
$
are also
independent of
\ytwoqubitresultphaseevolsin\
and yield
\begin{equation}
\label{eq-deriv_src_vs_mix-null}
\frac{d
\Ysssrcsigdisctimecentone
}{d \Ytwoqubitresultphaseevolsin}
=
0
\hspace{5mm} 
\mbox{and}
\hspace{5mm} 
\frac{d
\Ysssrcsigdisctimecenttwo
}{d \Ytwoqubitresultphaseevolsin}
=
0.
\end{equation}
The derivative
$
\frac{d \Ysssrcsigdisctimecentthree}
{d \Ytwoqubitresultphaseevolsin}
$
was already provided
above in\footnote{Due to the type of
notations employed throughout Section
\ref{sec-ml-previous},
we used a partial derivative notation in
(\ref{eq-deriv-sssrcsigdisctimecentthree-vs-twoqubitresultphaseevolsin}).
Anyway, the quantity considered in that equation 
(\ref{eq-deriv-sssrcsigdisctimecentthree-vs-twoqubitresultphaseevolsin})
is the same as
$
\frac{d \Ysssrcsigdisctimecentthree}
{d \Ytwoqubitresultphaseevolsin}
$
addressed here,
as explained in the previous footnote.}
(\ref{eq-deriv-sssrcsigdisctimecentthree-vs-twoqubitresultphaseevolsin}).
That derivative is used in
the term 
of (\ref{eq-cost-deriv-total})
related to the third source signal,
together with the {partial} derivative
$
\displaystyle
\frac{\partial
\Ymixfuncjacob
}
{\partial \Ysssrcsigdisctimecentthree}
$,
which is obtained
from
(\ref{eq-mixfuncjacobval})
and reads
\begin{equation}
\label{gradient-mixfuncjacob-vs-src-three}
\frac{\partial
\Ymixfuncjacob
}
{\partial \Ysssrcsigdisctimecentthree}
=
- 8
\Yparamqubitonestateplusmodulus ^2
\Yparamqubittwostateplusmodulus ^2
(
\Yparamqubittwostateplusmodulus ^2
-
\Yparamqubitonestateplusmodulus ^2
)
\sqrt{
1
-
\Yparamqubitonestateplusmodulus ^2
}
\sqrt{
1
-
\Yparamqubittwostateplusmodulus ^2
}
\sqrt{
1 -
\Ytwoqubitresultphaseevolsin ^2
}
\Ytwoqubitresultphaseevolsin
\sin
\Ytwoqubitresultphaseinit
.
\end{equation}

Inserting
(\ref{eq-mixfuncjacobval-gradient-partial}),
(\ref{eq-deriv_src_vs_mix-null}),
(\ref{gradient-mixfuncjacob-vs-src-three})
and
(\ref{eq-deriv-sssrcsigdisctimecentthree-vs-twoqubitresultphaseevolsin})
in
(\ref{eq-cost-deriv-total})
yields
\begin{eqnarray}
\label{eq-total-deriv-final}
\frac{
d \Ymixfuncjacob
}{
d \Ytwoqubitresultphaseevolsin
}
=
8
\Yparamqubitonestateplusmodulus ^2
\Yparamqubittwostateplusmodulus ^2
(
\Yparamqubittwostateplusmodulus ^2
-
\Yparamqubitonestateplusmodulus ^2
)
\sqrt{
1
-
\Yparamqubitonestateplusmodulus ^2
}
\sqrt{
1
-
\Yparamqubittwostateplusmodulus ^2
}
\frac
{
1 -
2 \Ytwoqubitresultphaseevolsin ^2
}
{
\sqrt{
1 -
\Ytwoqubitresultphaseevolsin ^2
}
}
\frac{1}
{\cos
\Ytwoqubitresultphaseinit}
-
8
\Yparamqubitonestateplusmodulus
\Yparamqubittwostateplusmodulus
(
\Yparamqubittwostateplusmodulus ^2
-
\Yparamqubitonestateplusmodulus ^2
) ^2
\Ytwoqubitresultphaseevolsin
\frac{\sin
\Ytwoqubitresultphaseinit}
{\cos
\Ytwoqubitresultphaseinit}
.
\end{eqnarray}

This completes the comment that we aimed at providing in this
document, concerning the total derivative to be used in the last term
of 
(\ref{eq-costlnmean-gradient-versiontwo}).

For the sake of clarity, we now conclude by providing the
explicit expression of the derivative
of the log-likelihood which
results from the 
complete
expression
(\ref{eq-total-deriv-final}).
Using this expression
together with
(\ref{eq-mixfuncjacobval}),
(\ref{eq-deriv-sssrcsigdisctimecentthree-vs-twoqubitresultphaseevolsin})
and
(\ref{eq-deriv_src_vs_mix-null})
allows us to rewrite
(\ref{eq-costlnmean-gradient-versiontwo})
as
\begin{eqnarray}
\nonumber
\frac{d \Ymlcostlnmean}
{d \Ytwoqubitresultphaseevolsin}
&
=
&
-
E_t \left[
\psi_{\Ytwoqubitresultphaseinit} (\Ytwoqubitresultphaseinit )
\left \{
\frac{
(
\Yparamqubittwostateplusmodulus ^2
-
\Yparamqubitonestateplusmodulus ^2
)
}
{
\Yparamqubitonestateplusmodulus
\Yparamqubittwostateplusmodulus
\sqrt{
1
-
\Yparamqubitonestateplusmodulus ^2
}
\sqrt{
1
-
\Yparamqubittwostateplusmodulus ^2
}
\sqrt{
1 -
\Ytwoqubitresultphaseevolsin ^2
}
\cos
\Ytwoqubitresultphaseinit
}
-
\frac
{
(1 -
2 \Ytwoqubitresultphaseevolsin ^2)
\sin
\Ytwoqubitresultphaseinit
}
{
(1 -
\Ytwoqubitresultphaseevolsin ^2
)
\Ytwoqubitresultphaseevolsin
\cos
\Ytwoqubitresultphaseinit
}
\right \}
\right]
\\
\label{eq-costlnmean-gradient-link-src-explicit}
&
&
-
E_t \left[
\frac
{
(1 -
2 \Ytwoqubitresultphaseevolsin ^2)
}
{
(1 -
\Ytwoqubitresultphaseevolsin ^2
)
\Ytwoqubitresultphaseevolsin
\cos ^2
\Ytwoqubitresultphaseinit
}
-
\frac{
(
\Yparamqubittwostateplusmodulus ^2
-
\Yparamqubitonestateplusmodulus ^2
)
\sin
\Ytwoqubitresultphaseinit}
{
\Yparamqubitonestateplusmodulus
\Yparamqubittwostateplusmodulus
\sqrt{
1
-
\Yparamqubitonestateplusmodulus ^2
}
\sqrt{
1
-
\Yparamqubittwostateplusmodulus ^2
}
\sqrt{
1 -
\Ytwoqubitresultphaseevolsin ^2
}
\cos ^2
\Ytwoqubitresultphaseinit
}
\right]
.
\end{eqnarray}

\end{document}

%% file: bss_com.tex
%
%
%



\newcommand{\ysssrcnb}
          {$\Ysssrcnb$}
\newcommand{\Ysssrcnb}
           {N}

\newcommand{\ysssensnb}
          {$\Ysssensnb$}
\newcommand{\Ysssensnb}
           {P}

\newcommand{\yssoutsepsystsigassocconttimecentnb}
{$\Yssoutsepsystsigassocconttimecentnb$}
\newcommand{\Yssoutsepsystsigassocconttimecentnb}
{L}

\newcommand{\yssbothtimevalnb}
          {$\Yssbothtimevalnb$}
\newcommand{\Yssbothtimevalnb}
           {L}

\newcommand{\yssbothfreqvalnb}
          {$\Yssbothfreqvalnb$}
\newcommand{\Yssbothfreqvalnb}
           {\Yssbothtimevalnb^{\prime}}

\newcommand{\yssbothscalevalnb}
          {$\Yssbothscalevalnb$}
\newcommand{\Yssbothscalevalnb}
           {\Yssbothtimevalnb^{\prime \prime}}



\newcommand{\yssconttimeval}
          {$\Yssconttimeval$}
\newcommand{\Yssconttimeval}
           {t}

\newcommand{\yssconttimevalsubone}
          {$\Yssconttimevalsubone$}
\newcommand{\Yssconttimevalsubone}
           {\Yssconttimeval _1}

\newcommand{\yssconttimevalsubtwo}
          {$\Yssconttimevalsubtwo$}
\newcommand{\Yssconttimevalsubtwo}
           {\Yssconttimeval _2}

\newcommand{\ysssubconttimeval}
          {$\Ysssubconttimeval$}
\newcommand{\Ysssubconttimeval}
           {p}

\newcommand{\yssconttimevalsubstd}
          {$\Yssconttimevalsubstd$}
\newcommand{\Yssconttimevalsubstd}
           {\Yssconttimeval _{\Ysssubconttimeval}}

\newcommand{\ysssubotherconttimeval}
          {$\Ysssubotherconttimeval$}
\newcommand{\Ysssubotherconttimeval}
           {q}

\newcommand{\yssconttimevalsubstdother}
          {$\Yssconttimevalsubstdother$}
\newcommand{\Yssconttimevalsubstdother}
           {\Yssconttimeval _{\Ysssubotherconttimeval}}

\newcommand{\yssconttimevalother}
          {$\Yssconttimevalother$}
\newcommand{\Yssconttimevalother}
           {t^{\prime}}

\newcommand{\yssconttimevalothersubone}
          {$\Yssconttimevalothersubone$}
\newcommand{\Yssconttimevalothersubone}
           {\Yssconttimevalother _1}

\newcommand{\yssconttimevalothersubtwo}
          {$\Yssconttimevalothersubtwo$}
\newcommand{\Yssconttimevalothersubtwo}
           {\Yssconttimevalother _2}





\newcommand{\yssdisctimeval}
          {$\Yssdisctimeval$}
\newcommand{\Yssdisctimeval}
           {n}

\newcommand{\yssdisctimevalsubone}
          {$\Yssdisctimevalsubone$}
\newcommand{\Yssdisctimevalsubone}
           {\Yssdisctimeval _1}

\newcommand{\yssdisctimevalsubtwo}
          {$\Yssdisctimevalsubtwo$}
\newcommand{\Yssdisctimevalsubtwo}
           {\Yssdisctimeval _2}

\newcommand{\ysssubdisctimeval}
          {$\Ysssubdisctimeval$}
\newcommand{\Ysssubdisctimeval}
           {i}

\newcommand{\yssdisctimevalsubstd}
          {$\Yssdisctimevalsubstd$}
\newcommand{\Yssdisctimevalsubstd}
           {\Yssdisctimeval _{\Ysssubdisctimeval}}

\newcommand{\ysssubotherdisctimeval}
          {$\Ysssubotherdisctimeval$}
\newcommand{\Ysssubotherdisctimeval}
           {j}

\newcommand{\yssdisctimevalsubstdother}
          {$\Yssdisctimevalsubstdother$}
\newcommand{\Yssdisctimevalsubstdother}
           {\Yssdisctimeval _{\Ysssubotherdisctimeval}}

\newcommand{\yssdisctimevalother}
          {$\Yssdisctimevalother$}
\newcommand{\Yssdisctimevalother}
           {m}

\newcommand{\yssdisctimevalothersubone}
          {$\Yssdisctimevalothersubone$}
\newcommand{\Yssdisctimevalothersubone}
           {\Yssdisctimevalother _1}

\newcommand{\yssdisctimevalothersubtwo}
          {$\Yssdisctimevalothersubtwo$}
\newcommand{\Yssdisctimevalothersubtwo}
           {\Yssdisctimevalother _2}


\newcommand{\yssdisctimelag}
          {$\Yssdisctimelag$}
\newcommand{\Yssdisctimelag}
           {l}

\newcommand{\yssdisctimelagsubone}
          {$\Yssdisctimelagsubone$}
\newcommand{\Yssdisctimelagsubone}
           {\Yssdisctimelag _1}

\newcommand{\yssdisctimelagsubtwo}
          {$\Yssdisctimelagsubtwo$}
\newcommand{\Yssdisctimelagsubtwo}
           {\Yssdisctimelag _2}


\newcommand{\ysscontfreqval}
          {$\Ysscontfreqval$}
\newcommand{\Ysscontfreqval}
           {\omega}

\newcommand{\ysssubcontfreqval}
          {$\Ysssubcontfreqval$}
\newcommand{\Ysssubcontfreqval}
           {\Ysssubconttimeval^{\prime}}

\newcommand{\ysscontfreqvalsubstd}
          {$\Ysscontfreqvalsubstd$}
\newcommand{\Ysscontfreqvalsubstd}
           {\Ysscontfreqval _{\Ysssubcontfreqval}}



\newcommand{\ysstimescalecontshiftval}
          {$\Ysstimescalecontshiftval$}
\newcommand{\Ysstimescalecontshiftval}
           {%
       \tau
       }

\newcommand{\ysstimescalecontscaleval}
          {$\Ysstimescalecontscaleval$}
\newcommand{\Ysstimescalecontscaleval}
           {%
       d
       }

\newcommand{\ysstimescalecontscaleexp}
          {$\Ysstimescalecontscaleexp$}
\newcommand{\Ysstimescalecontscaleexp}
           {%
       j
       }

\newcommand{\ysssubtimescalecontscaleval}
          {$\Ysssubtimescalecontscaleval$}
\newcommand{\Ysssubtimescalecontscaleval}
           {\Ysssubconttimeval^{\prime \prime}}

\newcommand{\ysstimescalecontscalevalsubstd}
          {$\Ysstimescalecontscalevalsubstd$}
\newcommand{\Ysstimescalecontscalevalsubstd}
           {\Ysstimescalecontscaleval _{\Ysssubtimescalecontscaleval}}

\newcommand{\ysstimescalecontmotherwav}
          {$\Ysstimescalecontmotherwav$}
\newcommand{\Ysstimescalecontmotherwav}
           {\psi}

\newcommand{\ysstimescalecontmotherwavval}
          {$\Ysstimescalecontmotherwavval$}
\newcommand{\Ysstimescalecontmotherwavval}
           {\Ysstimescalecontmotherwav ( \Yssconttimeval )}

\newcommand{\ysstimescalecontscaleshiftwav}
          {$\Ysstimescalecontscaleshiftwav$}
\newcommand{\Ysstimescalecontscaleshiftwav}
           {\Ysstimescalecontmotherwav
        _{\Ysstimescalecontshiftval , \Ysstimescalecontscaleval}
       }

\newcommand{\ysstimescalecontscaleshiftwavval}
          {$\Ysstimescalecontscaleshiftwavval$}
\newcommand{\Ysstimescalecontscaleshiftwavval}
           {\Ysstimescalecontscaleshiftwav ( \Yssconttimeval )}

\newcommand{\ysstimescalecontcoefnot}
          {$\Ysstimescalecontcoefnot$}
\newcommand{\Ysstimescalecontcoefnot}
           {W}


\newcommand{\ysssepsystlag}
          {$\Ysssepsystlag$}
\newcommand{\Ysssepsystlag}
           {k}

\newcommand{\ysscomplfiltlag}
          {$\Ysscomplfiltlag$}
\newcommand{\Ysscomplfiltlag}
           {m}

\newcommand{\ysscomplfiltlagother}
          {$\Ysscomplfiltlagother$}
\newcommand{\Ysscomplfiltlagother}
           {l}


\newcommand{\ysswayindex}
          {$\Ysswayindex$}
\newcommand{\Ysswayindex}
           {i}

\newcommand{\ysswayindexother}
          {$\Ysswayindexother$}
\newcommand{\Ysswayindexother}
           {j}

\newcommand{\ysswayindexthird}
          {$\Ysswayindexthird$}
\newcommand{\Ysswayindexthird}
           {k}

\newcommand{\ysswayindexfourth}
          {$\Ysswayindexfourth$}
\newcommand{\Ysswayindexfourth}
           {l}




\newcommand{\ysssrcsiginnovdisctimecentvec}
          {$\Ysssrcsiginnovdisctimecentvec$}
\newcommand{\Ysssrcsiginnovdisctimecentvec}
           {p}

\newcommand{\ysssrcsiginnovdisctimecentvecval}
          {$\Ysssrcsiginnovdisctimecentvecval$}
\newcommand{\Ysssrcsiginnovdisctimecentvecval}
           {\Ysssrcsiginnovdisctimecentvec (\Yssdisctimeval)}

\newcommand{\ysssrcsiginnnovdisctimecentone}
          {$\Ysssrcsiginnovdisctimecentone$}
\newcommand{\Ysssrcsiginnovdisctimecentone}
           {\Ysssrcsiginnovdisctimecentvec _{1}}

\newcommand{\ysssrcsiginnovdisctimecentoneval}
          {$\Ysssrcsiginnovdisctimecentoneval$}
\newcommand{\Ysssrcsiginnovdisctimecentoneval}
           {\Ysssrcsiginnovdisctimecentone (\Yssdisctimeval)}

\newcommand{\ysssrcsiginnnovdisctimecenttwo}
          {$\Ysssrcsiginnovdisctimecenttwo$}
\newcommand{\Ysssrcsiginnovdisctimecenttwo}
           {\Ysssrcsiginnovdisctimecentvec _{2}}

\newcommand{\ysssrcsiginnovdisctimecenttwoval}
          {$\Ysssrcsiginnovdisctimecenttwoval$}
\newcommand{\Ysssrcsiginnovdisctimecenttwoval}
           {\Ysssrcsiginnovdisctimecenttwo (\Yssdisctimeval)}

\newcommand{\ysssrcsiginnovdisctimecentindex}
          {$\Ysssrcsiginnovdisctimecentindex$}
\newcommand{\Ysssrcsiginnovdisctimecentindex}
           {\Ysssrcsiginnovdisctimecentvec _{\Ysswayindex}}

\newcommand{\ysssrcsiginnovdisctimecentindexval}
          {$\Ysssrcsiginnovdisctimecentindexval$}
\newcommand{\Ysssrcsiginnovdisctimecentindexval}
           {\Ysssrcsiginnovdisctimecentindex (\Yssdisctimeval)}

\newcommand{\ysssrcsiginnovdisctimecentindexother}
          {$\Ysssrcsiginnovdisctimecentindexother$}
\newcommand{\Ysssrcsiginnovdisctimecentindexother}
           {\Ysssrcsiginnovdisctimecentvec _{\Ysswayindexother}}

\newcommand{\ysssrcsiginnovdisctimecentindexotherval}
          {$\Ysssrcsiginnovdisctimecentindexotherval$}
\newcommand{\Ysssrcsiginnovdisctimecentindexotherval}
           {\Ysssrcsiginnovdisctimecentindexother (\Yssdisctimeval)}


\newcommand{\ysssrcsiginnovdisctimecentveczt}
          {$\Ysssrcsiginnovdisctimecentveczt$}
\newcommand{\Ysssrcsiginnovdisctimecentveczt}
           {P}

\newcommand{\ysssrcsiginnovdisctimecentvecztval}
          {$\Ysssrcsiginnovdisctimecentvecztval$}
\newcommand{\Ysssrcsiginnovdisctimecentvecztval}
           {\Ysssrcsiginnovdisctimecentveczt (z)}

\newcommand{\ysssrcsiginnovdisctimecentindexzt}
          {$\Ysssrcsiginnovdisctimecentindexzt$}
\newcommand{\Ysssrcsiginnovdisctimecentindexzt}
           {\Ysssrcsiginnovdisctimecentveczt _{\Ysswayindex}}

\newcommand{\ysssrcsiginnovdisctimecentindexztval}
          {$\Ysssrcsiginnovdisctimecentindexztval$}
\newcommand{\Ysssrcsiginnovdisctimecentindexztval}
           {\Ysssrcsiginnovdisctimecentindexzt (z)}



\newcommand{\ysssrcsigconttimecentvec}
          {$\Ysssrcsigconttimecentvec$}
\newcommand{\Ysssrcsigconttimecentvec}
           {s}

\newcommand{\ysssrcsigconttimecentvecval}
          {$\Ysssrcsigconttimecentvecval$}
\newcommand{\Ysssrcsigconttimecentvecval}
           {\Ysssrcsigconttimecentvec (\Yssconttimeval)}

\newcommand{\ysssrcsigconttimecentone}
          {$\Ysssrcsigconttimecentone$}
\newcommand{\Ysssrcsigconttimecentone}
           {\Ysssrcsigconttimecentvec _{1}}

\newcommand{\ysssrcsigconttimecentoneval}
          {$\Ysssrcsigconttimecentoneval$}
\newcommand{\Ysssrcsigconttimecentoneval}
           {\Ysssrcsigconttimecentone (\Yssconttimeval)}

\newcommand{\ysssrcsigconttimecenttwo}
          {$\Ysssrcsigconttimecenttwo$}
\newcommand{\Ysssrcsigconttimecenttwo}
           {\Ysssrcsigconttimecentvec _{2}}

\newcommand{\ysssrcsigconttimecenttwoval}
          {$\Ysssrcsigconttimecenttwoval$}
\newcommand{\Ysssrcsigconttimecenttwoval}
           {\Ysssrcsigconttimecenttwo (\Yssconttimeval)}

\newcommand{\ysssrcsigconttimecentthree}
          {$\Ysssrcsigconttimecentthree$}
\newcommand{\Ysssrcsigconttimecentthree}
           {\Ysssrcsigconttimecentvec _{3}}

\newcommand{\ysssrcsigconttimecentthreeval}
          {$\Ysssrcsigconttimecentthreeval$}
\newcommand{\Ysssrcsigconttimecentthreeval}
           {\Ysssrcsigconttimecentthree (\Yssconttimeval)}

\newcommand{\ysssrcsigconttimecentindex}
          {$\Ysssrcsigconttimecentindex$}
\newcommand{\Ysssrcsigconttimecentindex}
           {\Ysssrcsigconttimecentvec _{\Ysswayindex}}

\newcommand{\ysssrcsigconttimecentindexval}
          {$\Ysssrcsigconttimecentindexval$}
\newcommand{\Ysssrcsigconttimecentindexval}
           {\Ysssrcsigconttimecentindex (\Yssconttimeval)}

\newcommand{\ysssrcsigconttimecentindexother}
          {$\Ysssrcsigconttimecentindexother$}
\newcommand{\Ysssrcsigconttimecentindexother}
           {\Ysssrcsigconttimecentvec _{\Ysswayindexother}}

\newcommand{\ysssrcsigconttimecentindexotherval}
          {$\Ysssrcsigconttimecentindexotherval$}
\newcommand{\Ysssrcsigconttimecentindexotherval}
           {\Ysssrcsigconttimecentindexother (\Yssconttimeval)}

\newcommand{\ysssrcsigconttimecentindexthird}
          {$\Ysssrcsigconttimecentindexthird$}
\newcommand{\Ysssrcsigconttimecentindexthird}
           {\Ysssrcsigconttimecentvec _{\Ysswayindexthird}}

\newcommand{\ysssrcsigconttimecentindexthirdval}
          {$\Ysssrcsigconttimecentindexthirdval$}
\newcommand{\Ysssrcsigconttimecentindexthirdval}
           {\Ysssrcsigconttimecentindexthird (\Yssconttimeval)}

\newcommand{\ysssrcsigconttimecentindexfourth}
          {$\Ysssrcsigconttimecentindexfourth$}
\newcommand{\Ysssrcsigconttimecentindexfourth}
           {\Ysssrcsigconttimecentvec _{\Ysswayindexfourth}}

\newcommand{\ysssrcsigconttimecentindexfourthval}
          {$\Ysssrcsigconttimecentindexfourthval$}
\newcommand{\Ysssrcsigconttimecentindexfourthval}
           {\Ysssrcsigconttimecentindexfourth (\Yssconttimeval)}

\newcommand{\ysssrcsigconttimecentlast}
          {$\Ysssrcsigconttimecentlast$}
\newcommand{\Ysssrcsigconttimecentlast}
           {\Ysssrcsigconttimecentvec _{\Ysssrcnb}}

\newcommand{\ysssrcsigconttimecentlastval}
          {$\Ysssrcsigconttimecentlastval$}
\newcommand{\Ysssrcsigconttimecentlastval}
           {\Ysssrcsigconttimecentlast (\Yssconttimeval)}


\newcommand{\ysssrcsigassocconttimecentvec}
          {$\Ysssrcsigassocconttimecentvec$}
\newcommand{\Ysssrcsigassocconttimecentvec}
           {\Ysssrcsigconttimecentvec ^{\prime}}

\newcommand{\ysssrcsigassocconttimecentvecval}
          {$\Ysssrcsigassocconttimecentvecval$}
\newcommand{\Ysssrcsigassocconttimecentvecval}
           {\Ysssrcsigassocconttimecentvec (\Yssconttimeval)}

\newcommand{\ysssrcsigassocconttimecentone}
          {$\Ysssrcsigassocconttimecentone$}
\newcommand{\Ysssrcsigassocconttimecentone}
           {\Ysssrcsigassocconttimecentvec _{1}}

\newcommand{\ysssrcsigassocconttimecentoneval}
          {$\Ysssrcsigassocconttimecentoneval$}
\newcommand{\Ysssrcsigassocconttimecentoneval}
           {\Ysssrcsigassocconttimecentone (\Yssconttimeval)}

\newcommand{\ysssrcsigassocconttimecenttwo}
          {$\Ysssrcsigassocconttimecenttwo$}
\newcommand{\Ysssrcsigassocconttimecenttwo}
           {\Ysssrcsigassocconttimecentvec _{2}}

\newcommand{\ysssrcsigassocconttimecenttwoval}
          {$\Ysssrcsigassocconttimecenttwoval$}
\newcommand{\Ysssrcsigassocconttimecenttwoval}
           {\Ysssrcsigassocconttimecenttwo (\Yssconttimeval)}

\newcommand{\ysssrcsigassocconttimecentthree}
          {$\Ysssrcsigassocconttimecentthree$}
\newcommand{\Ysssrcsigassocconttimecentthree}
           {\Ysssrcsigassocconttimecentvec _{3}}

\newcommand{\ysssrcsigassocconttimecentthreeval}
          {$\Ysssrcsigassocconttimecentthreeval$}
\newcommand{\Ysssrcsigassocconttimecentthreeval}
           {\Ysssrcsigassocconttimecentthree (\Yssconttimeval)}

\newcommand{\ysssrcsigassocconttimecentindex}
          {$\Ysssrcsigassocconttimecentindex$}
\newcommand{\Ysssrcsigassocconttimecentindex}
           {\Ysssrcsigassocconttimecentvec _{\Ysswayindex}}

\newcommand{\ysssrcsigassocconttimecentindexval}
          {$\Ysssrcsigassocconttimecentindexval$}
\newcommand{\Ysssrcsigassocconttimecentindexval}
           {\Ysssrcsigassocconttimecentindex (\Yssconttimeval)}

\newcommand{\ysssrcsigassocconttimecentindexother}
          {$\Ysssrcsigassocconttimecentindexother$}
\newcommand{\Ysssrcsigassocconttimecentindexother}
           {\Ysssrcsigassocconttimecentvec _{\Ysswayindexother}}

\newcommand{\ysssrcsigassocconttimecentindexotherval}
          {$\Ysssrcsigassocconttimecentindexotherval$}
\newcommand{\Ysssrcsigassocconttimecentindexotherval}
           {\Ysssrcsigassocconttimecentindexother (\Yssconttimeval)}

\newcommand{\ysssrcsigassocconttimecentindexthird}
          {$\Ysssrcsigassocconttimecentindexthird$}
\newcommand{\Ysssrcsigassocconttimecentindexthird}
           {\Ysssrcsigassocconttimecentvec _{\Ysswayindexthird}}

\newcommand{\ysssrcsigassocconttimecentindexthirdval}
          {$\Ysssrcsigassocconttimecentindexthirdval$}
\newcommand{\Ysssrcsigassocconttimecentindexthirdval}
           {\Ysssrcsigassocconttimecentindexthird (\Yssconttimeval)}

\newcommand{\ysssrcsigassocconttimecentindexfourth}
{$\Ysssrcsigassocconttimecentindexfourth$}
\newcommand{\Ysssrcsigassocconttimecentindexfourth}
{\Ysssrcsigassocconttimecentvec _{\Ysswayindexfourth}}

\newcommand{\ysssrcsigassocconttimecentindexfourthval}
{$\Ysssrcsigassocconttimecentindexfourthval$}
\newcommand{\Ysssrcsigassocconttimecentindexfourthval}
{\Ysssrcsigassocconttimecentindexfourth (\Yssconttimeval)}

\newcommand{\ysssrcsigassocconttimecentlast}
          {$\Ysssrcsigassocconttimecentlast$}
\newcommand{\Ysssrcsigassocconttimecentlast}
           {\Ysssrcsigassocconttimecentvec _{\Ysssrcnb}}

\newcommand{\ysssrcsigassocconttimecentlastval}
          {$\Ysssrcsigassocconttimecentlastval$}
\newcommand{\Ysssrcsigassocconttimecentlastval}
           {\Ysssrcsigassocconttimecentlast (\Yssconttimeval)}


\newcommand{\ysssrcsigconttimecontfreqnoncentvec}
          {$\Ysssrcsigconttimecontfreqnoncentvec$}
\newcommand{\Ysssrcsigconttimecontfreqnoncentvec}
           {S}

\newcommand{\ysssrcsigconttimecontfreqnoncentvecval}
          {$\Ysssrcsigconttimecontfreqnoncentvecval$}
\newcommand{\Ysssrcsigconttimecontfreqnoncentvecval}
           {\Ysssrcsigconttimecontfreqnoncentvec ( \Yssconttimeval , \Ysscontfreqval ) }

\newcommand{\ysssrcsigconttimecontfreqnoncentone}
          {$\Ysssrcsigconttimecontfreqnoncentone$}
\newcommand{\Ysssrcsigconttimecontfreqnoncentone}
           {\Ysssrcsigconttimecontfreqnoncentvec _{1}}

\newcommand{\ysssrcsigconttimecontfreqnoncentoneval}
          {$\Ysssrcsigconttimecontfreqnoncentoneval$}
\newcommand{\Ysssrcsigconttimecontfreqnoncentoneval}
           {\Ysssrcsigconttimecontfreqnoncentone ( \Yssconttimeval , \Ysscontfreqval ) }

\newcommand{\ysssrcsigconttimecontfreqnoncenttwo}
          {$\Ysssrcsigconttimecontfreqnoncenttwo$}
\newcommand{\Ysssrcsigconttimecontfreqnoncenttwo}
           {\Ysssrcsigconttimecontfreqnoncentvec _{2}}

\newcommand{\ysssrcsigconttimecontfreqnoncenttwoval}
          {$\Ysssrcsigconttimecontfreqnoncenttwoval$}
\newcommand{\Ysssrcsigconttimecontfreqnoncenttwoval}
           {\Ysssrcsigconttimecontfreqnoncenttwo ( \Yssconttimeval , \Ysscontfreqval ) }

\newcommand{\ysssrcsigconttimecontfreqnoncentindex}
          {$\Ysssrcsigconttimecontfreqnoncentindex$}
\newcommand{\Ysssrcsigconttimecontfreqnoncentindex}
           {\Ysssrcsigconttimecontfreqnoncentvec _{\Ysswayindex}}

\newcommand{\ysssrcsigconttimecontfreqnoncentindexval}
          {$\Ysssrcsigconttimecontfreqnoncentindexval$}
\newcommand{\Ysssrcsigconttimecontfreqnoncentindexval}
           {\Ysssrcsigconttimecontfreqnoncentindex ( \Yssconttimeval , \Ysscontfreqval ) }

\newcommand{\ysssrcsigconttimecontfreqnoncentindexother}
          {$\Ysssrcsigconttimecontfreqnoncentindexother$}
\newcommand{\Ysssrcsigconttimecontfreqnoncentindexother}
           {\Ysssrcsigconttimecontfreqnoncentvec _{\Ysswayindexother}}

\newcommand{\ysssrcsigconttimecontfreqnoncentindexotherval}
          {$\Ysssrcsigconttimecontfreqnoncentindexotherval$}
\newcommand{\Ysssrcsigconttimecontfreqnoncentindexotherval}
           {\Ysssrcsigconttimecontfreqnoncentindexother ( \Yssconttimeval , \Ysscontfreqval ) }

\newcommand{\ysssrcsigconttimecontfreqnoncentindexthird}
          {$\Ysssrcsigconttimecontfreqnoncentindexthird$}
\newcommand{\Ysssrcsigconttimecontfreqnoncentindexthird}
           {\Ysssrcsigconttimecontfreqnoncentvec _{\Ysswayindexthird}}

\newcommand{\ysssrcsigconttimecontfreqnoncentindexthirdval}
          {$\Ysssrcsigconttimecontfreqnoncentindexthirdval$}
\newcommand{\Ysssrcsigconttimecontfreqnoncentindexthirdval}
           {\Ysssrcsigconttimecontfreqnoncentindexthird ( \Yssconttimeval , \Ysscontfreqval ) }

\newcommand{\ysssrcsigconttimecontfreqnoncentindexfourth}
          {$\Ysssrcsigconttimecontfreqnoncentindexfourth$}
\newcommand{\Ysssrcsigconttimecontfreqnoncentindexfourth}
           {\Ysssrcsigconttimecontfreqnoncentvec _{\Ysswayindexfourth}}

\newcommand{\ysssrcsigconttimecontfreqnoncentindexfourthval}
          {$\Ysssrcsigconttimecontfreqnoncentindexfourthval$}
\newcommand{\Ysssrcsigconttimecontfreqnoncentindexfourthval}
           {\Ysssrcsigconttimecontfreqnoncentindexfourth ( \Yssconttimeval , \Ysscontfreqval ) }


\newcommand{\ysssrcsigconttimecontscalecentindexother}
          {$\Ysssrcsigconttimecontscalecentindexother$}
\newcommand{\Ysssrcsigconttimecontscalecentindexother}
           {\Ysstimescalecontcoefnot _{\Ysssrcsigconttimecentvec _{\Ysswayindexother}}}

\newcommand{\ysssrcsigconttimecontscalecentindexotherval}
          {$\Ysssrcsigconttimecontscalecentindexotherval$}
\newcommand{\Ysssrcsigconttimecontscalecentindexotherval}
           {\Ysssrcsigconttimecontscalecentindexother
        ( \Ysstimescalecontshiftval , \Ysstimescalecontscaleval )
           }

\newcommand{\ysssrcsigconttimecontscalecentindexothervalsubstd}
          {$\Ysssrcsigconttimecontscalecentindexothervalsubstd$}
\newcommand{\Ysssrcsigconttimecontscalecentindexothervalsubstd}
           {\Ysssrcsigconttimecontscalecentindexother
        ( \Ysstimescalecontshiftval , \Ysstimescalecontscalevalsubstd )
           }


\newcommand{\ysssrcsigdisctimecentvec}
          {$\Ysssrcsigdisctimecentvec$}
\newcommand{\Ysssrcsigdisctimecentvec}
           {s}

\newcommand{\ysssrcsigdisctimecentvecval}
          {$\Ysssrcsigdisctimecentvecval$}
\newcommand{\Ysssrcsigdisctimecentvecval}
           {\Ysssrcsigdisctimecentvec (\Yssdisctimeval)}

\newcommand{\ysssrcsigdisctimecentone}
          {$\Ysssrcsigdisctimecentone$}
\newcommand{\Ysssrcsigdisctimecentone}
           {\Ysssrcsigdisctimecentvec _{1}}

\newcommand{\ysssrcsigdisctimecentoneval}
          {$\Ysssrcsigdisctimecentoneval$}
\newcommand{\Ysssrcsigdisctimecentoneval}
           {\Ysssrcsigdisctimecentone (\Yssdisctimeval)}

\newcommand{\ysssrcsigdisctimecenttwo}
          {$\Ysssrcsigdisctimecenttwo$}
\newcommand{\Ysssrcsigdisctimecenttwo}
           {\Ysssrcsigdisctimecentvec _{2}}

\newcommand{\ysssrcsigdisctimecenttwoval}
          {$\Ysssrcsigdisctimecenttwoval$}
\newcommand{\Ysssrcsigdisctimecenttwoval}
           {\Ysssrcsigdisctimecenttwo (\Yssdisctimeval)}

\newcommand{\ysssrcsigdisctimecentthree}
          {$\Ysssrcsigdisctimecentthree$}
\newcommand{\Ysssrcsigdisctimecentthree}
           {\Ysssrcsigdisctimecentvec _{3}}

\newcommand{\ysssrcsigdisctimecentthreeval}
          {$\Ysssrcsigdisctimecentthreeval$}
\newcommand{\Ysssrcsigdisctimecentthreeval}
           {\Ysssrcsigdisctimecentthree (\Yssdisctimeval)}

\newcommand{\ysssrcsigdisctimecentindex}
          {$\Ysssrcsigdisctimecentindex$}
\newcommand{\Ysssrcsigdisctimecentindex}
           {\Ysssrcsigdisctimecentvec _{\Ysswayindex}}

\newcommand{\ysssrcsigdisctimecentindexval}
          {$\Ysssrcsigdisctimecentindexval$}
\newcommand{\Ysssrcsigdisctimecentindexval}
           {\Ysssrcsigdisctimecentindex (\Yssdisctimeval)}

\newcommand{\ysssrcsigdisctimecentindexother}
          {$\Ysssrcsigdisctimecentindexother$}
\newcommand{\Ysssrcsigdisctimecentindexother}
           {\Ysssrcsigdisctimecentvec _{\Ysswayindexother}}

\newcommand{\ysssrcsigdisctimecentindexotherval}
          {$\Ysssrcsigdisctimecentindexotherval$}
\newcommand{\Ysssrcsigdisctimecentindexotherval}
           {\Ysssrcsigdisctimecentindexother (\Yssdisctimeval)}

\newcommand{\ysssrcsigdisctimecentindexthird}
          {$\Ysssrcsigdisctimecentindexthird$}
\newcommand{\Ysssrcsigdisctimecentindexthird}
           {\Ysssrcsigdisctimecentvec _{\Ysswayindexthird}}

\newcommand{\ysssrcsigdisctimecentindexthirdval}
          {$\Ysssrcsigdisctimecentindexthirdval$}
\newcommand{\Ysssrcsigdisctimecentindexthirdval}
           {\Ysssrcsigdisctimecentindexthird (\Yssdisctimeval)}

\newcommand{\ysssrcsigdisctimecentlast}
          {$\Ysssrcsigdisctimecentlast$}
\newcommand{\Ysssrcsigdisctimecentlast}
           {\Ysssrcsigdisctimecentvec _{\Ysssrcnb}}

\newcommand{\ysssrcsigdisctimecentlastval}
          {$\Ysssrcsigdisctimecentlastval$}
\newcommand{\Ysssrcsigdisctimecentlastval}
           {\Ysssrcsigdisctimecentlast (\Yssdisctimeval)}


\newcommand{\ysssrcsigassoctwodisctimecentvec}
          {$\Ysssrcsigassoctwodisctimecentvec$}
\newcommand{\Ysssrcsigassoctwodisctimecentvec}
           {\tilde{\Ysssrcsigdisctimecentvec}}

\newcommand{\ysssrcsigassoctwodisctimecentvecval}
          {$\Ysssrcsigassoctwodisctimecentvecval$}
\newcommand{\Ysssrcsigassoctwodisctimecentvecval}
           {\Ysssrcsigassoctwodisctimecentvec (\Yssdisctimeval)}

\newcommand{\ysssrcsigassoctwodisctimecentone}
{$\Ysssrcsigassoctwodisctimecentone$}
\newcommand{\Ysssrcsigassoctwodisctimecentone}
{\Ysssrcsigassoctwodisctimecentvec _{1}}

\newcommand{\ysssrcsigassoctwodisctimecentoneval}
{$\Ysssrcsigassoctwodisctimecentoneval$}
\newcommand{\Ysssrcsigassoctwodisctimecentoneval}
{\Ysssrcsigassoctwodisctimecentone (\Yssdisctimeval)}

\newcommand{\ysssrcsigassoctwodisctimecenttwo}
{$\Ysssrcsigassoctwodisctimecenttwo$}
\newcommand{\Ysssrcsigassoctwodisctimecenttwo}
{\Ysssrcsigassoctwodisctimecentvec _{2}}

\newcommand{\ysssrcsigassoctwodisctimecenttwoval}
{$\Ysssrcsigassoctwodisctimecenttwoval$}
\newcommand{\Ysssrcsigassoctwodisctimecenttwoval}
{\Ysssrcsigassoctwodisctimecenttwo (\Yssdisctimeval)}

\newcommand{\ysssrcsigassoctwodisctimecentindex}
{$\Ysssrcsigassoctwodisctimecentindex$}
\newcommand{\Ysssrcsigassoctwodisctimecentindex}
{\Ysssrcsigassoctwodisctimecentvec _{\Ysswayindex}}

\newcommand{\ysssrcsigassoctwodisctimecentindexval}
{$\Ysssrcsigassoctwodisctimecentindexval$}
\newcommand{\Ysssrcsigassoctwodisctimecentindexval}
{\Ysssrcsigassoctwodisctimecentindex (\Yssdisctimeval)}

\newcommand{\ysssrcsigassoctwodisctimecentindexother}
          {$\Ysssrcsigassoctwodisctimecentindexother$}
\newcommand{\Ysssrcsigassoctwodisctimecentindexother}
           {\Ysssrcsigassoctwodisctimecentvec _{\Ysswayindexother}}

\newcommand{\ysssrcsigassoctwodisctimecentindexotherval}
          {$\Ysssrcsigassoctwodisctimecentindexotherval$}
\newcommand{\Ysssrcsigassoctwodisctimecentindexotherval}
           {\Ysssrcsigassoctwodisctimecentindexother (\Yssdisctimeval)}


\newcommand{\ysssrcsigdisctimecontfreqnotonenoncentvec}
          {$\Ysssrcsigdisctimecontfreqnotonenoncentvec$}
\newcommand{\Ysssrcsigdisctimecontfreqnotonenoncentvec}
           {S}

\newcommand{\ysssrcsigdisctimecontfreqnotonenoncentvecval}
          {$\Ysssrcsigdisctimecontfreqnotonenoncentvecval$}
\newcommand{\Ysssrcsigdisctimecontfreqnotonenoncentvecval}
           {\Ysssrcsigdisctimecontfreqnotonenoncentvec ( \Ysscontfreqval ) }

\newcommand{\ysssrcsigdisctimecontfreqnotonenoncentone}
          {$\Ysssrcsigdisctimecontfreqnotonenoncentone$}
\newcommand{\Ysssrcsigdisctimecontfreqnotonenoncentone}
           {\Ysssrcsigdisctimecontfreqnotonenoncentvec _{1}}

\newcommand{\ysssrcsigdisctimecontfreqnotonenoncentoneval}
          {$\Ysssrcsigdisctimecontfreqnotonenoncentoneval$}
\newcommand{\Ysssrcsigdisctimecontfreqnotonenoncentoneval}
           {\Ysssrcsigdisctimecontfreqnotonenoncentone (\Ysscontfreqval ) }

\newcommand{\ysssrcsigdisctimecontfreqnotonenoncentindexother}
          {$\Ysssrcsigdisctimecontfreqnotonenoncentindexother$}
\newcommand{\Ysssrcsigdisctimecontfreqnotonenoncentindexother}
           {\Ysssrcsigdisctimecontfreqnotonenoncentvec _{\Ysswayindexother}}

\newcommand{\ysssrcsigdisctimecontfreqnotonenoncentindexotherval}
          {$\Ysssrcsigdisctimecontfreqnotonenoncentindexotherval$}
\newcommand{\Ysssrcsigdisctimecontfreqnotonenoncentindexotherval}
           {\Ysssrcsigdisctimecontfreqnotonenoncentindexother (\Ysscontfreqval ) }

\newcommand{\ysssrcsigdisctimecontfreqnotonenoncentlast}
          {$\Ysssrcsigdisctimecontfreqnotonenoncentlast$}
\newcommand{\Ysssrcsigdisctimecontfreqnotonenoncentlast}
           {\Ysssrcsigdisctimecontfreqnotonenoncentvec _{\Ysssrcnb}}

\newcommand{\ysssrcsigdisctimecontfreqnotonenoncentlastval}
          {$\Ysssrcsigdisctimecontfreqnotonenoncentlastval$}
\newcommand{\Ysssrcsigdisctimecontfreqnotonenoncentlastval}
           {\Ysssrcsigdisctimecontfreqnotonenoncentlast (\Ysscontfreqval ) }


\newcommand{\ysssrcsigassocdisctimecontfreqnotonenoncentvec}
          {$\Ysssrcsigassocdisctimecontfreqnotonenoncentvec$}
\newcommand{\Ysssrcsigassocdisctimecontfreqnotonenoncentvec}
           {\Ysssrcsigdisctimecontfreqnotonenoncentvec ^{\prime}}

\newcommand{\ysssrcsigassocdisctimecontfreqnotonenoncentvecval}
          {$\Ysssrcsigassocdisctimecontfreqnotonenoncentvecval$}
\newcommand{\Ysssrcsigassocdisctimecontfreqnotonenoncentvecval}
           {\Ysssrcsigassocdisctimecontfreqnotonenoncentvec ( \Ysscontfreqval ) }

\newcommand{\ysssrcsigassocdisctimecontfreqnotonenoncentone}
          {$\Ysssrcsigassocdisctimecontfreqnotonenoncentone$}
\newcommand{\Ysssrcsigassocdisctimecontfreqnotonenoncentone}
           {\Ysssrcsigassocdisctimecontfreqnotonenoncentvec _{1}}

\newcommand{\ysssrcsigassocdisctimecontfreqnotonenoncentoneval}
          {$\Ysssrcsigassocdisctimecontfreqnotonenoncentoneval$}
\newcommand{\Ysssrcsigassocdisctimecontfreqnotonenoncentoneval}
           {\Ysssrcsigassocdisctimecontfreqnotonenoncentone (\Ysscontfreqval ) }

\newcommand{\ysssrcsigassocdisctimecontfreqnotonenoncentindexother}
          {$\Ysssrcsigassocdisctimecontfreqnotonenoncentindexother$}
\newcommand{\Ysssrcsigassocdisctimecontfreqnotonenoncentindexother}
           {\Ysssrcsigassocdisctimecontfreqnotonenoncentvec _{\Ysswayindexother}}

\newcommand{\ysssrcsigassocdisctimecontfreqnotonenoncentindexotherval}
          {$\Ysssrcsigassocdisctimecontfreqnotonenoncentindexotherval$}
\newcommand{\Ysssrcsigassocdisctimecontfreqnotonenoncentindexotherval}
           {\Ysssrcsigassocdisctimecontfreqnotonenoncentindexother (\Ysscontfreqval ) }

\newcommand{\ysssrcsigassocdisctimecontfreqnotonenoncentlast}
          {$\Ysssrcsigassocdisctimecontfreqnotonenoncentlast$}
\newcommand{\Ysssrcsigassocdisctimecontfreqnotonenoncentlast}
           {\Ysssrcsigassocdisctimecontfreqnotonenoncentvec _{\Ysssrcnb}}

\newcommand{\ysssrcsigassocdisctimecontfreqnotonenoncentlastval}
          {$\Ysssrcsigassocdisctimecontfreqnotonenoncentlastval$}
\newcommand{\Ysssrcsigassocdisctimecontfreqnotonenoncentlastval}
           {\Ysssrcsigassocdisctimecontfreqnotonenoncentlast (\Ysscontfreqval ) }


\newcommand{\ysssrcsigdisctimecontfreqnoncentvec}
          {$\Ysssrcsigdisctimecontfreqnoncentvec$}
\newcommand{\Ysssrcsigdisctimecontfreqnoncentvec}
           {S}

\newcommand{\ysssrcsigdisctimecontfreqnoncentvecval}
          {$\Ysssrcsigdisctimecontfreqnoncentvecval$}
\newcommand{\Ysssrcsigdisctimecontfreqnoncentvecval}
           {\Ysssrcsigdisctimecontfreqnoncentvec ( \Yssdisctimeval , \Ysscontfreqval ) }

\newcommand{\ysssrcsigdisctimecontfreqnoncentone}
          {$\Ysssrcsigdisctimecontfreqnoncentone$}
\newcommand{\Ysssrcsigdisctimecontfreqnoncentone}
           {\Ysssrcsigdisctimecontfreqnoncentvec _{1}}

\newcommand{\ysssrcsigdisctimecontfreqnoncentoneval}
          {$\Ysssrcsigdisctimecontfreqnoncentoneval$}
\newcommand{\Ysssrcsigdisctimecontfreqnoncentoneval}
           {\Ysssrcsigdisctimecontfreqnoncentone ( \Yssdisctimeval , \Ysscontfreqval ) }

\newcommand{\ysssrcsigdisctimecontfreqnoncentindex}
          {$\Ysssrcsigdisctimecontfreqnoncentindex$}
\newcommand{\Ysssrcsigdisctimecontfreqnoncentindex}
           {\Ysssrcsigdisctimecontfreqnoncentvec _{\Ysswayindex}}

\newcommand{\ysssrcsigdisctimecontfreqnoncentindexval}
          {$\Ysssrcsigdisctimecontfreqnoncentindexval$}
\newcommand{\Ysssrcsigdisctimecontfreqnoncentindexval}
           {\Ysssrcsigdisctimecontfreqnoncentindex ( \Yssdisctimeval , \Ysscontfreqval ) }

\newcommand{\ysssrcsigdisctimecontfreqnoncentindexother}
          {$\Ysssrcsigdisctimecontfreqnoncentindexother$}
\newcommand{\Ysssrcsigdisctimecontfreqnoncentindexother}
           {\Ysssrcsigdisctimecontfreqnoncentvec _{\Ysswayindexother}}

\newcommand{\ysssrcsigdisctimecontfreqnoncentindexotherval}
          {$\Ysssrcsigdisctimecontfreqnoncentindexotherval$}
\newcommand{\Ysssrcsigdisctimecontfreqnoncentindexotherval}
           {\Ysssrcsigdisctimecontfreqnoncentindexother ( \Yssdisctimeval , \Ysscontfreqval ) }

\newcommand{\ysssrcsigdisctimecontfreqnoncentindexthird}
          {$\Ysssrcsigdisctimecontfreqnoncentindexthird$}
\newcommand{\Ysssrcsigdisctimecontfreqnoncentindexthird}
           {\Ysssrcsigdisctimecontfreqnoncentvec _{\Ysswayindexthird}}

\newcommand{\ysssrcsigdisctimecontfreqnoncentindexthirdval}
          {$\Ysssrcsigdisctimecontfreqnoncentindexthirdval$}
\newcommand{\Ysssrcsigdisctimecontfreqnoncentindexthirdval}
           {\Ysssrcsigdisctimecontfreqnoncentindexthird ( \Yssdisctimeval , \Ysscontfreqval ) }


\newcommand{\ysssrcsigassoctwodisctimecontfreqnoncentvec}
          {$\Ysssrcsigassoctwodisctimecontfreqnoncentvec$}
\newcommand{\Ysssrcsigassoctwodisctimecontfreqnoncentvec}
           {\tilde{\Ysssrcsigdisctimecontfreqnoncentvec}}

\newcommand{\ysssrcsigassoctwodisctimecontfreqnoncentvecval}
          {$\Ysssrcsigassoctwodisctimecontfreqnoncentvecval$}
\newcommand{\Ysssrcsigassoctwodisctimecontfreqnoncentvecval}
           {\Ysssrcsigassoctwodisctimecontfreqnoncentvec ( \Yssdisctimeval , \Ysscontfreqval ) }

\newcommand{\ysssrcsigassoctwodisctimecontfreqnoncentindexother}
          {$\Ysssrcsigassoctwodisctimecontfreqnoncentindexother$}
\newcommand{\Ysssrcsigassoctwodisctimecontfreqnoncentindexother}
           {\Ysssrcsigassoctwodisctimecontfreqnoncentvec _{\Ysswayindexother}}

\newcommand{\ysssrcsigassoctwodisctimecontfreqnoncentindexotherval}
          {$\Ysssrcsigassoctwodisctimecontfreqnoncentindexotherval$}
\newcommand{\Ysssrcsigassoctwodisctimecontfreqnoncentindexotherval}
           {\Ysssrcsigassoctwodisctimecontfreqnoncentindexother ( \Yssdisctimeval , \Ysscontfreqval ) }


\newcommand{\ysssrcsigdisctimecentveczt}
          {$\Ysssrcsigdisctimecentveczt$}
\newcommand{\Ysssrcsigdisctimecentveczt}
           {S}

\newcommand{\ysssrcsigdisctimecentvecztval}
          {$\Ysssrcsigdisctimecentvecztval$}
\newcommand{\Ysssrcsigdisctimecentvecztval}
           {\Ysssrcsigdisctimecentveczt (z)}

\newcommand{\ysssrcsigdisctimecentonezt}
          {$\Ysssrcsigdisctimecentonezt$}
\newcommand{\Ysssrcsigdisctimecentonezt}
           {\Ysssrcsigdisctimecentveczt _{1}}

\newcommand{\ysssrcsigdisctimecentoneztval}
          {$\Ysssrcsigdisctimecentoneztval$}
\newcommand{\Ysssrcsigdisctimecentoneztval}
           {\Ysssrcsigdisctimecentonezt (z)}

\newcommand{\ysssrcsigdisctimecenttwozt}
          {$\Ysssrcsigdisctimecenttwozt$}
\newcommand{\Ysssrcsigdisctimecenttwozt}
           {\Ysssrcsigdisctimecentveczt _{2}}

\newcommand{\ysssrcsigdisctimecenttwoztval}
          {$\Ysssrcsigdisctimecenttwoztval$}
\newcommand{\Ysssrcsigdisctimecenttwoztval}
           {\Ysssrcsigdisctimecenttwozt (z)}

\newcommand{\ysssrcsigdisctimecentthreezt}
          {$\Ysssrcsigdisctimecentthreezt$}
\newcommand{\Ysssrcsigdisctimecentthreezt}
           {\Ysssrcsigdisctimecentveczt _{3}}

\newcommand{\ysssrcsigdisctimecentthreeztval}
          {$\Ysssrcsigdisctimecentthreeztval$}
\newcommand{\Ysssrcsigdisctimecentthreeztval}
           {\Ysssrcsigdisctimecentthreezt (z)}

\newcommand{\ysssrcsigdisctimecentindexzt}
          {$\Ysssrcsigdisctimecentindexzt$}
\newcommand{\Ysssrcsigdisctimecentindexzt}
           {\Ysssrcsigdisctimecentveczt _{\Ysswayindex}}

\newcommand{\ysssrcsigdisctimecentindexztval}
          {$\Ysssrcsigdisctimecentindexztval$}
\newcommand{\Ysssrcsigdisctimecentindexztval}
           {\Ysssrcsigdisctimecentindexzt (z)}

\newcommand{\ysssrcsigdisctimecentindexotherzt}
          {$\Ysssrcsigdisctimecentindexotherzt$}
\newcommand{\Ysssrcsigdisctimecentindexotherzt}
           {\Ysssrcsigdisctimecentveczt _{\Ysswayindexother}}

\newcommand{\ysssrcsigdisctimecentindexotherztval}
          {$\Ysssrcsigdisctimecentindexotherztval$}
\newcommand{\Ysssrcsigdisctimecentindexotherztval}
           {\Ysssrcsigdisctimecentindexotherzt (z)}

\newcommand{\ysssrcsigdisctimecentlastzt}
          {$\Ysssrcsigdisctimecentlastzt$}
\newcommand{\Ysssrcsigdisctimecentlastzt}
           {\Ysssrcsigdisctimecentveczt _{\Ysssrcnb}}

\newcommand{\ysssrcsigdisctimecentlastztval}
          {$\Ysssrcsigdisctimecentlastztval$}
\newcommand{\Ysssrcsigdisctimecentlastztval}
           {\Ysssrcsigdisctimecentlastzt (z)}



\newcommand{\yssmixsigconttimecentvec}
          {$\Yssmixsigconttimecentvec$}
\newcommand{\Yssmixsigconttimecentvec}
           {x}

\newcommand{\yssmixsigconttimecentvecval}
          {$\Yssmixsigconttimecentvecval$}
\newcommand{\Yssmixsigconttimecentvecval}
           {\Yssmixsigconttimecentvec (\Yssconttimeval)}

\newcommand{\yssmixsigconttimecentone}
          {$\Yssmixsigconttimecentone$}
\newcommand{\Yssmixsigconttimecentone}
           {\Yssmixsigconttimecentvec _{1}}

\newcommand{\yssmixsigconttimecentoneval}
          {$\Yssmixsigconttimecentoneval$}
\newcommand{\Yssmixsigconttimecentoneval}
           {\Yssmixsigconttimecentone (\Yssconttimeval)}

\newcommand{\yssmixsigconttimecenttwo}
          {$\Yssmixsigconttimecenttwo$}
\newcommand{\Yssmixsigconttimecenttwo}
           {\Yssmixsigconttimecentvec _{2}}

\newcommand{\yssmixsigconttimecenttwoval}
          {$\Yssmixsigconttimecenttwoval$}
\newcommand{\Yssmixsigconttimecenttwoval}
           {\Yssmixsigconttimecenttwo (\Yssconttimeval)}

\newcommand{\yssmixsigconttimecentindex}
          {$\Yssmixsigconttimecentindex$}
\newcommand{\Yssmixsigconttimecentindex}
           {\Yssmixsigconttimecentvec _{\Ysswayindex}}

\newcommand{\yssmixsigconttimecentindexval}
          {$\Yssmixsigconttimecentindexval$}
\newcommand{\Yssmixsigconttimecentindexval}
           {\Yssmixsigconttimecentindex (\Yssconttimeval)}

\newcommand{\yssmixsigconttimecentindexother}
          {$\Yssmixsigconttimecentindexother$}
\newcommand{\Yssmixsigconttimecentindexother}
           {\Yssmixsigconttimecentvec _{\Ysswayindexother}}

\newcommand{\yssmixsigconttimecentindexotherval}
          {$\Yssmixsigconttimecentindexotherval$}
\newcommand{\Yssmixsigconttimecentindexotherval}
           {\Yssmixsigconttimecentindexother (\Yssconttimeval)}

\newcommand{\yssmixsigconttimecentindexthird}
          {$\Yssmixsigconttimecentindexthird$}
\newcommand{\Yssmixsigconttimecentindexthird}
           {\Yssmixsigconttimecentvec _{\Ysswayindexthird}}

\newcommand{\yssmixsigconttimecentindexthirdval}
          {$\Yssmixsigconttimecentindexthirdval$}
\newcommand{\Yssmixsigconttimecentindexthirdval}
           {\Yssmixsigconttimecentindexthird (\Yssconttimeval)}

\newcommand{\yssmixsigconttimecentlast}
          {$\Yssmixsigconttimecentlast$}
\newcommand{\Yssmixsigconttimecentlast}
           {\Yssmixsigconttimecentvec _{\Ysssensnb}}

\newcommand{\yssmixsigconttimecentlastval}
          {$\Yssmixsigconttimecentlastval$}
\newcommand{\Yssmixsigconttimecentlastval}
           {\Yssmixsigconttimecentlast (\Yssconttimeval)}

\newcommand{\yssmixsigconttimecentsrcnb}
          {$\Yssmixsigconttimecentsrcnb$}
\newcommand{\Yssmixsigconttimecentsrcnb}
           {\Yssmixsigconttimecentvec _{\Ysssrcnb}}


\newcommand{\yssmixsigconttimecentsrcnbval}
          {$\Yssmixsigconttimecentsrcnbval$}
\newcommand{\Yssmixsigconttimecentsrcnbval}
           {\Yssmixsigconttimecentsrcnb (\Yssconttimeval)}


\newcommand{\yssmixsigassocconttimecentvec}
{$\Yssmixsigassocconttimecentvec$}
\newcommand{\Yssmixsigassocconttimecentvec}
{\Yssmixsigconttimecentvec ^{\prime}}

\newcommand{\yssmixsigassocconttimecentvecval}
{$\Yssmixsigassocconttimecentvecval$}
\newcommand{\Yssmixsigassocconttimecentvecval}
{\Yssmixsigassocconttimecentvec (\Yssconttimeval)}

\newcommand{\yssmixsigassocconttimecentone}
{$\Yssmixsigassocconttimecentone$}
\newcommand{\Yssmixsigassocconttimecentone}
{\Yssmixsigassocconttimecentvec _{1}}

\newcommand{\yssmixsigassocconttimecentoneval}
{$\Yssmixsigassocconttimecentoneval$}
\newcommand{\Yssmixsigassocconttimecentoneval}
{\Yssmixsigassocconttimecentone (\Yssconttimeval)}

\newcommand{\yssmixsigassocconttimecentindex}
{$\Yssmixsigassocconttimecentindex$}
\newcommand{\Yssmixsigassocconttimecentindex}
{\Yssmixsigassocconttimecentvec _{\Ysswayindex}}

\newcommand{\yssmixsigassocconttimecentindexval}
{$\Yssmixsigassocconttimecentindexval$}
\newcommand{\Yssmixsigassocconttimecentindexval}
{\Yssmixsigassocconttimecentindex (\Yssconttimeval)}

\newcommand{\yssmixsigassocconttimecentindexthird}
{$\Yssmixsigassocconttimecentindexthird$}
\newcommand{\Yssmixsigassocconttimecentindexthird}
{\Yssmixsigassocconttimecentvec _{\Ysswayindexthird}}

\newcommand{\yssmixsigassocconttimecentindexthirdval}
{$\Yssmixsigassocconttimecentindexthirdval$}
\newcommand{\Yssmixsigassocconttimecentindexthirdval}
{\Yssmixsigassocconttimecentindexthird (\Yssconttimeval)}


\newcommand{\yssmixsigconttimecontfreqnoncentvec}
          {$\Yssmixsigconttimecontfreqnoncentvec$}
\newcommand{\Yssmixsigconttimecontfreqnoncentvec}
           {X}

\newcommand{\yssmixsigconttimecontfreqnoncentvecval}
          {$\Yssmixsigconttimecontfreqnoncentvecval$}
\newcommand{\Yssmixsigconttimecontfreqnoncentvecval}
           {\Yssmixsigconttimecontfreqnoncentvec ( \Yssconttimeval , \Ysscontfreqval ) }

\newcommand{\yssmixsigconttimecontfreqnoncentone}
          {$\Yssmixsigconttimecontfreqnoncentone$}
\newcommand{\Yssmixsigconttimecontfreqnoncentone}
           {\Yssmixsigconttimecontfreqnoncentvec _{1}}

\newcommand{\yssmixsigconttimecontfreqnoncentoneval}
          {$\Yssmixsigconttimecontfreqnoncentoneval$}
\newcommand{\Yssmixsigconttimecontfreqnoncentoneval}
           {\Yssmixsigconttimecontfreqnoncentone ( \Yssconttimeval , \Ysscontfreqval ) }

\newcommand{\yssmixsigconttimecontfreqnoncenttwo}
          {$\Yssmixsigconttimecontfreqnoncenttwo$}
\newcommand{\Yssmixsigconttimecontfreqnoncenttwo}
           {\Yssmixsigconttimecontfreqnoncentvec _{2}}

\newcommand{\yssmixsigconttimecontfreqnoncenttwoval}
          {$\Yssmixsigconttimecontfreqnoncenttwoval$}
\newcommand{\Yssmixsigconttimecontfreqnoncenttwoval}
           {\Yssmixsigconttimecontfreqnoncenttwo ( \Yssconttimeval , \Ysscontfreqval ) }

\newcommand{\yssmixsigconttimecontfreqnoncentindex}
          {$\Yssmixsigconttimecontfreqnoncentindex$}
\newcommand{\Yssmixsigconttimecontfreqnoncentindex}
           {\Yssmixsigconttimecontfreqnoncentvec _{\Ysswayindex}}

\newcommand{\yssmixsigconttimecontfreqnoncentindexval}
          {$\Yssmixsigconttimecontfreqnoncentindexval$}
\newcommand{\Yssmixsigconttimecontfreqnoncentindexval}
           {\Yssmixsigconttimecontfreqnoncentindex ( \Yssconttimeval , \Ysscontfreqval ) }

\newcommand{\yssmixsigconttimecontfreqnoncentindexother}
          {$\Yssmixsigconttimecontfreqnoncentindexother$}
\newcommand{\Yssmixsigconttimecontfreqnoncentindexother}
           {\Yssmixsigconttimecontfreqnoncentvec _{\Ysswayindexother}}

\newcommand{\yssmixsigconttimecontfreqnoncentindexotherval}
          {$\Yssmixsigconttimecontfreqnoncentindexotherval$}
\newcommand{\Yssmixsigconttimecontfreqnoncentindexotherval}
           {\Yssmixsigconttimecontfreqnoncentindexother ( \Yssconttimeval , \Ysscontfreqval ) }

\newcommand{\yssmixsigconttimecontfreqnoncentindexthird}
          {$\Yssmixsigconttimecontfreqnoncentindexthird$}
\newcommand{\Yssmixsigconttimecontfreqnoncentindexthird}
           {\Yssmixsigconttimecontfreqnoncentvec _{\Ysswayindexthird}}

\newcommand{\yssmixsigconttimecontfreqnoncentindexthirdval}
          {$\Yssmixsigconttimecontfreqnoncentindexthirdval$}
\newcommand{\Yssmixsigconttimecontfreqnoncentindexthirdval}
           {\Yssmixsigconttimecontfreqnoncentindexthird ( \Yssconttimeval , \Ysscontfreqval ) }


\newcommand{\yssmixsigconttimecontscalecentindex}
          {$\Yssmixsigconttimecontscalecentindex$}
\newcommand{\Yssmixsigconttimecontscalecentindex}
           {\Ysstimescalecontcoefnot _{\Yssmixsigconttimecentvec _{\Ysswayindex}}}

\newcommand{\yssmixsigconttimecontscalecentindexval}
          {$\Yssmixsigconttimecontscalecentindexval$}
\newcommand{\Yssmixsigconttimecontscalecentindexval}
           {\Yssmixsigconttimecontscalecentindex
        ( \Ysstimescalecontshiftval , \Ysstimescalecontscaleval )
           }

\newcommand{\yssmixsigconttimecontscalecentindexvalsubstd}
          {$\Yssmixsigconttimecontscalecentindexvalsubstd$}
\newcommand{\Yssmixsigconttimecontscalecentindexvalsubstd}
           {\Yssmixsigconttimecontscalecentindex
        ( \Ysstimescalecontshiftval , \Ysstimescalecontscalevalsubstd )
           }


\newcommand{\yssmixsigdisctimecentvec}
          {$\Yssmixsigdisctimecentvec$}
\newcommand{\Yssmixsigdisctimecentvec}
           {x}

\newcommand{\yssmixsigdisctimecentvecval}
          {$\Yssmixsigdisctimecentvecval$}
\newcommand{\Yssmixsigdisctimecentvecval}
           {\Yssmixsigdisctimecentvec (\Yssdisctimeval)}

\newcommand{\yssmixsigdisctimecentone}
          {$\Yssmixsigdisctimecentone$}
\newcommand{\Yssmixsigdisctimecentone}
           {\Yssmixsigdisctimecentvec _{1}}

\newcommand{\yssmixsigdisctimecentoneval}
          {$\Yssmixsigdisctimecentoneval$}
\newcommand{\Yssmixsigdisctimecentoneval}
           {\Yssmixsigdisctimecentone (\Yssdisctimeval)}

\newcommand{\yssmixsigdisctimecenttwo}
          {$\Yssmixsigdisctimecenttwo$}
\newcommand{\Yssmixsigdisctimecenttwo}
           {\Yssmixsigdisctimecentvec _{2}}

\newcommand{\yssmixsigdisctimecenttwoval}
          {$\Yssmixsigdisctimecenttwoval$}
\newcommand{\Yssmixsigdisctimecenttwoval}
           {\Yssmixsigdisctimecenttwo (\Yssdisctimeval)}

\newcommand{\yssmixsigdisctimecentthree}
{$\Yssmixsigdisctimecentthree$}
\newcommand{\Yssmixsigdisctimecentthree}
{\Yssmixsigdisctimecentvec _{3}}

\newcommand{\yssmixsigdisctimecentthreeval}
{$\Yssmixsigdisctimecentthreeval$}
\newcommand{\Yssmixsigdisctimecentthreeval}
{\Yssmixsigdisctimecentthree (\Yssdisctimeval)}

\newcommand{\yssmixsigdisctimecentindex}
          {$\Yssmixsigdisctimecentindex$}
\newcommand{\Yssmixsigdisctimecentindex}
           {\Yssmixsigdisctimecentvec _{\Ysswayindex}}

\newcommand{\yssmixsigdisctimecentindexval}
          {$\Yssmixsigdisctimecentindexval$}
\newcommand{\Yssmixsigdisctimecentindexval}
           {\Yssmixsigdisctimecentindex (\Yssdisctimeval)}

\newcommand{\yssmixsigdisctimecentindexother}
{$\Yssmixsigdisctimecentindexother$}
\newcommand{\Yssmixsigdisctimecentindexother}
{\Yssmixsigdisctimecentvec _{\Ysswayindexother}}

\newcommand{\yssmixsigdisctimecentindexotherval}
{$\Yssmixsigdisctimecentindexotherval$}
\newcommand{\Yssmixsigdisctimecentindexotherval}
{\Yssmixsigdisctimecentindexother (\Yssdisctimeval)}

\newcommand{\yssmixsigdisctimecentlast}
          {$\Yssmixsigdisctimecentlast$}
\newcommand{\Yssmixsigdisctimecentlast}
           {\Yssmixsigdisctimecentvec _{\Ysssensnb}}

\newcommand{\yssmixsigdisctimecentlastval}
          {$\Yssmixsigdisctimecentlastval$}
\newcommand{\Yssmixsigdisctimecentlastval}
           {\Yssmixsigdisctimecentlast (\Yssdisctimeval)}

\newcommand{\yssmixsigdisctimecentsrcnb}
          {$\Yssmixsigdisctimecentsrcnb$}
\newcommand{\Yssmixsigdisctimecentsrcnb}
           {\Yssmixsigdisctimecentvec _{\Ysssrcnb}}


\newcommand{\yssmixsigdisctimecentsrcnbval}
          {$\Yssmixsigdisctimecentsrcnbval$}
\newcommand{\Yssmixsigdisctimecentsrcnbval}
           {\Yssmixsigdisctimecentsrcnb (\Yssdisctimeval)}


\newcommand{\yssmixsigdisctimecontfreqnotonenoncentvec}
          {$\Yssmixsigdisctimecontfreqnotonenoncentvec$}
\newcommand{\Yssmixsigdisctimecontfreqnotonenoncentvec}
           {X}

\newcommand{\yssmixsigdisctimecontfreqnotonenoncentvecval}
          {$\Yssmixsigdisctimecontfreqnotonenoncentvecval$}
\newcommand{\Yssmixsigdisctimecontfreqnotonenoncentvecval}
           {\Yssmixsigdisctimecontfreqnotonenoncentvec ( \Ysscontfreqval ) }

\newcommand{\yssmixsigdisctimecontfreqnotonenoncentone}
          {$\Yssmixsigdisctimecontfreqnotonenoncentone$}
\newcommand{\Yssmixsigdisctimecontfreqnotonenoncentone}
           {\Yssmixsigdisctimecontfreqnotonenoncentvec _{1}}

\newcommand{\yssmixsigdisctimecontfreqnotonenoncentoneval}
          {$\Yssmixsigdisctimecontfreqnotonenoncentoneval$}
\newcommand{\Yssmixsigdisctimecontfreqnotonenoncentoneval}
           {\Yssmixsigdisctimecontfreqnotonenoncentone (\Ysscontfreqval ) }

\newcommand{\yssmixsigdisctimecontfreqnotonenoncenttwo}
          {$\Yssmixsigdisctimecontfreqnotonenoncenttwo$}
\newcommand{\Yssmixsigdisctimecontfreqnotonenoncenttwo}
           {\Yssmixsigdisctimecontfreqnotonenoncentvec _{2}}

\newcommand{\yssmixsigdisctimecontfreqnotonenoncenttwoval}
          {$\Yssmixsigdisctimecontfreqnotonenoncenttwoval$}
\newcommand{\Yssmixsigdisctimecontfreqnotonenoncenttwoval}
           {\Yssmixsigdisctimecontfreqnotonenoncenttwo (\Ysscontfreqval ) }

\newcommand{\yssmixsigdisctimecontfreqnotonenoncentindex}
          {$\Yssmixsigdisctimecontfreqnotonenoncentindex$}
\newcommand{\Yssmixsigdisctimecontfreqnotonenoncentindex}
           {\Yssmixsigdisctimecontfreqnotonenoncentvec _{\Ysswayindex}}

\newcommand{\yssmixsigdisctimecontfreqnotonenoncentindexval}
          {$\Yssmixsigdisctimecontfreqnotonenoncentindexval$}
\newcommand{\Yssmixsigdisctimecontfreqnotonenoncentindexval}
           {\Yssmixsigdisctimecontfreqnotonenoncentindex (\Ysscontfreqval ) }

\newcommand{\yssmixsigdisctimecontfreqnotonenoncentsrcnb}
          {$\Yssmixsigdisctimecontfreqnotonenoncentsrcnb$}
\newcommand{\Yssmixsigdisctimecontfreqnotonenoncentsrcnb}
           {\Yssmixsigdisctimecontfreqnotonenoncentvec _{\Ysssrcnb}}

\newcommand{\yssmixsigdisctimecontfreqnotonenoncentsrcnbval}
          {$\Yssmixsigdisctimecontfreqnotonenoncentsrcnbval$}
\newcommand{\Yssmixsigdisctimecontfreqnotonenoncentsrcnbval}
           {\Yssmixsigdisctimecontfreqnotonenoncentsrcnb (\Ysscontfreqval ) }


\newcommand{\yssmixsigdisctimecontfreqnoncentvec}
          {$\Yssmixsigdisctimecontfreqnoncentvec$}
\newcommand{\Yssmixsigdisctimecontfreqnoncentvec}
           {X}

\newcommand{\yssmixsigdisctimecontfreqnoncentvecval}
          {$\Yssmixsigdisctimecontfreqnoncentvecval$}
\newcommand{\Yssmixsigdisctimecontfreqnoncentvecval}
           {\Yssmixsigdisctimecontfreqnoncentvec ( \Yssdisctimeval , \Ysscontfreqval ) }

\newcommand{\yssmixsigdisctimecontfreqnoncentone}
          {$\Yssmixsigdisctimecontfreqnoncentone$}
\newcommand{\Yssmixsigdisctimecontfreqnoncentone}
           {\Yssmixsigdisctimecontfreqnoncentvec _{1}}

\newcommand{\yssmixsigdisctimecontfreqnoncentoneval}
          {$\Yssmixsigdisctimecontfreqnoncentoneval$}
\newcommand{\Yssmixsigdisctimecontfreqnoncentoneval}
           {\Yssmixsigdisctimecontfreqnoncentone ( \Yssdisctimeval , \Ysscontfreqval ) }

\newcommand{\yssmixsigdisctimecontfreqnoncentindex}
          {$\Yssmixsigdisctimecontfreqnoncentindex$}
\newcommand{\Yssmixsigdisctimecontfreqnoncentindex}
           {\Yssmixsigdisctimecontfreqnoncentvec _{\Ysswayindex}}

\newcommand{\yssmixsigdisctimecontfreqnoncentindexval}
          {$\Yssmixsigdisctimecontfreqnoncentindexval$}
\newcommand{\Yssmixsigdisctimecontfreqnoncentindexval}
           {\Yssmixsigdisctimecontfreqnoncentindex ( \Yssdisctimeval , \Ysscontfreqval ) }

\newcommand{\yssmixsigdisctimecontfreqnoncentindexother}
          {$\Yssmixsigdisctimecontfreqnoncentindexother$}
\newcommand{\Yssmixsigdisctimecontfreqnoncentindexother}
           {\Yssmixsigdisctimecontfreqnoncentvec _{\Ysswayindexother}}

\newcommand{\yssmixsigdisctimecontfreqnoncentindexotherval}
          {$\Yssmixsigdisctimecontfreqnoncentindexotherval$}
\newcommand{\Yssmixsigdisctimecontfreqnoncentindexotherval}
           {\Yssmixsigdisctimecontfreqnoncentindexother ( \Yssdisctimeval , \Ysscontfreqval ) }


\newcommand{\yssmixsigdisctimecentveczt}
          {$\Yssmixsigdisctimecentveczt$}
\newcommand{\Yssmixsigdisctimecentveczt}
           {X}

\newcommand{\yssmixsigdisctimecentvecztval}
          {$\Yssmixsigdisctimecentvecztval$}
\newcommand{\Yssmixsigdisctimecentvecztval}
           {\Yssmixsigdisctimecentveczt (z)}

\newcommand{\yssmixsigdisctimecentonezt}
          {$\Yssmixsigdisctimecentonezt$}
\newcommand{\Yssmixsigdisctimecentonezt}
           {\Yssmixsigdisctimecentveczt _{1}}

\newcommand{\yssmixsigdisctimecentoneztval}
          {$\Yssmixsigdisctimecentoneztval$}
\newcommand{\Yssmixsigdisctimecentoneztval}
           {\Yssmixsigdisctimecentonezt (z)}

\newcommand{\yssmixsigdisctimecenttwozt}
          {$\Yssmixsigdisctimecenttwozt$}
\newcommand{\Yssmixsigdisctimecenttwozt}
           {\Yssmixsigdisctimecentveczt _{2}}

\newcommand{\yssmixsigdisctimecenttwoztval}
          {$\Yssmixsigdisctimecenttwoztval$}
\newcommand{\Yssmixsigdisctimecenttwoztval}
           {\Yssmixsigdisctimecenttwozt (z)}

\newcommand{\yssmixsigdisctimecentindexzt}
          {$\Yssmixsigdisctimecentindexzt$}
\newcommand{\Yssmixsigdisctimecentindexzt}
           {\Yssmixsigdisctimecentveczt _{\Ysswayindex}}

\newcommand{\yssmixsigdisctimecentindexztval}
          {$\Yssmixsigdisctimecentindexztval$}
\newcommand{\Yssmixsigdisctimecentindexztval}
           {\Yssmixsigdisctimecentindexzt (z)}

\newcommand{\yssmixsigdisctimecentlastzt}
          {$\Yssmixsigdisctimecentlastzt$}
\newcommand{\Yssmixsigdisctimecentlastzt}
           {\Yssmixsigdisctimecentveczt _{\Ysssensnb}}

\newcommand{\yssmixsigdisctimecentlastztval}
          {$\Yssmixsigdisctimecentlastztval$}
\newcommand{\Yssmixsigdisctimecentlastztval}
           {\Yssmixsigdisctimecentlastzt (z)}



\newcommand{\yssintermsepsystsigdisctimecentvec}
          {$\Yssintermsepsystsigdisctimecentvec$}
\newcommand{\Yssintermsepsystsigdisctimecentvec}
           {u}

\newcommand{\yssintermsepsystsigdisctimecentone}
          {$\Yssintermsepsystsigdisctimecentone$}
\newcommand{\Yssintermsepsystsigdisctimecentone}
           {\Yssintermsepsystsigdisctimecentvec _{1}}

\newcommand{\yssintermsepsystsigdisctimecentoneval}
          {$\Yssintermsepsystsigdisctimecentoneval$}
\newcommand{\Yssintermsepsystsigdisctimecentoneval}
           {\Yssintermsepsystsigdisctimecentone ( \Yssdisctimeval )}

\newcommand{\yssintermsepsystsigdisctimecenttwo}
          {$\Yssintermsepsystsigdisctimecenttwo$}
\newcommand{\Yssintermsepsystsigdisctimecenttwo}
           {\Yssintermsepsystsigdisctimecentvec _{2}}

\newcommand{\yssintermsepsystsigdisctimecenttwoval}
          {$\Yssintermsepsystsigdisctimecenttwoval$}
\newcommand{\Yssintermsepsystsigdisctimecenttwoval}
           {\Yssintermsepsystsigdisctimecenttwo ( \Yssdisctimeval )}

\newcommand{\yssintermsepsystsigdisctimecentindex}
          {$\Yssintermsepsystsigdisctimecentindex$}
\newcommand{\Yssintermsepsystsigdisctimecentindex}
           {\Yssintermsepsystsigdisctimecentvec _{\Ysswayindex}}

\newcommand{\yssintermsepsystsigdisctimecentindexval}
          {$\Yssintermsepsystsigdisctimecentindexval$}
\newcommand{\Yssintermsepsystsigdisctimecentindexval}
           {\Yssintermsepsystsigdisctimecentindex ( \Yssdisctimeval )}

\newcommand{\yssintermsepsystsigdisctimecentindexother}
          {$\Yssintermsepsystsigdisctimecentindexother$}
\newcommand{\Yssintermsepsystsigdisctimecentindexother}
           {\Yssintermsepsystsigdisctimecentvec _{\Ysswayindexother}}


\newcommand{\yssoutsepsystsigconttimecentvec}
          {$\Yssoutsepsystsigconttimecentvec$}
\newcommand{\Yssoutsepsystsigconttimecentvec}
           {y}

\newcommand{\yssoutsepsystsigconttimecentvecval}
          {$\Yssoutsepsystsigconttimecentvecval$}
\newcommand{\Yssoutsepsystsigconttimecentvecval}
           {\Yssoutsepsystsigconttimecentvec (\Yssconttimeval)}

\newcommand{\yssoutsepsystsigconttimecentone}
          {$\Yssoutsepsystsigconttimecentone$}
\newcommand{\Yssoutsepsystsigconttimecentone}
           {\Yssoutsepsystsigconttimecentvec _{1}}

\newcommand{\yssoutsepsystsigconttimecentoneval}
          {$\Yssoutsepsystsigconttimecentoneval$}
\newcommand{\Yssoutsepsystsigconttimecentoneval}
           {\Yssoutsepsystsigconttimecentone (\Yssconttimeval)}

\newcommand{\yssoutsepsystsigconttimecentindex}
          {$\Yssoutsepsystsigconttimecentindex$}
\newcommand{\Yssoutsepsystsigconttimecentindex}
           {\Yssoutsepsystsigconttimecentvec _{\Ysswayindex}}

\newcommand{\yssoutsepsystsigconttimecentindexval}
          {$\Yssoutsepsystsigconttimecentindexval$}
\newcommand{\Yssoutsepsystsigconttimecentindexval}
           {\Yssoutsepsystsigconttimecentindex (\Yssconttimeval)}

\newcommand{\yssoutsepsystsigconttimecentindexother}
          {$\Yssoutsepsystsigconttimecentindexother$}
\newcommand{\Yssoutsepsystsigconttimecentindexother}
           {\Yssoutsepsystsigconttimecentvec _{\Ysswayindexother}}

\newcommand{\yssoutsepsystsigconttimecentindexotherval}
          {$\Yssoutsepsystsigconttimecentindexotherval$}
\newcommand{\Yssoutsepsystsigconttimecentindexotherval}
           {\Yssoutsepsystsigconttimecentindexother (\Yssconttimeval)}

\newcommand{\yssoutsepsystsigconttimecentindexfourth}
{$\Yssoutsepsystsigconttimecentindexfourth$}
\newcommand{\Yssoutsepsystsigconttimecentindexfourth}
{\Yssoutsepsystsigconttimecentvec _{\Ysswayindexfourth}}

\newcommand{\yssoutsepsystsigconttimecentindexfourthval}
{$\Yssoutsepsystsigconttimecentindexfourthval$}
\newcommand{\Yssoutsepsystsigconttimecentindexfourthval}
{\Yssoutsepsystsigconttimecentindexfourth (\Yssconttimeval)}


\newcommand{\yssoutsepsystsigassocconttimecentvec}
          {$\Yssoutsepsystsigassocconttimecentvec$}
\newcommand{\Yssoutsepsystsigassocconttimecentvec}
           {\Yssoutsepsystsigconttimecentvec ^{\prime}}

\newcommand{\yssoutsepsystsigassocconttimecentvecval}
          {$\Yssoutsepsystsigassocconttimecentvecval$}
\newcommand{\Yssoutsepsystsigassocconttimecentvecval}
           {\Yssoutsepsystsigassocconttimecentvec (\Yssconttimeval)}

\newcommand{\yssoutsepsystsigassocconttimecentone}
          {$\Yssoutsepsystsigassocconttimecentone$}
\newcommand{\Yssoutsepsystsigassocconttimecentone}
           {\Yssoutsepsystsigassocconttimecentvec _{1}}

\newcommand{\yssoutsepsystsigassocconttimecentoneval}
          {$\Yssoutsepsystsigassocconttimecentoneval$}
\newcommand{\Yssoutsepsystsigassocconttimecentoneval}
           {\Yssoutsepsystsigassocconttimecentone (\Yssconttimeval)}

\newcommand{\yssoutsepsystsigassocconttimecentindex}
          {$\Yssoutsepsystsigassocconttimecentindex$}
\newcommand{\Yssoutsepsystsigassocconttimecentindex}
           {\Yssoutsepsystsigassocconttimecentvec _{\Ysswayindex}}

\newcommand{\yssoutsepsystsigassocconttimecentindexval}
          {$\Yssoutsepsystsigassocconttimecentindexval$}
\newcommand{\Yssoutsepsystsigassocconttimecentindexval}
           {\Yssoutsepsystsigassocconttimecentindex (\Yssconttimeval)}

\newcommand{\yssoutsepsystsigassocconttimecentindexother}
          {$\Yssoutsepsystsigassocconttimecentindexother$}
\newcommand{\Yssoutsepsystsigassocconttimecentindexother}
           {\Yssoutsepsystsigassocconttimecentvec _{\Ysswayindexother}}

\newcommand{\yssoutsepsystsigassocconttimecentindexotherval}
          {$\Yssoutsepsystsigassocconttimecentindexotherval$}
\newcommand{\Yssoutsepsystsigassocconttimecentindexotherval}
           {\Yssoutsepsystsigassocconttimecentindexother (\Yssconttimeval)}

\newcommand{\yssoutsepsystsigassocconttimecentindexfourth}
{$\Yssoutsepsystsigassocconttimecentindexfourth$}
\newcommand{\Yssoutsepsystsigassocconttimecentindexfourth}
{\Yssoutsepsystsigassocconttimecentvec _{\Ysswayindexfourth}}

\newcommand{\yssoutsepsystsigassocconttimecentindexfourthval}
{$\Yssoutsepsystsigassocconttimecentindexfourthval$}
\newcommand{\Yssoutsepsystsigassocconttimecentindexfourthval}
{\Yssoutsepsystsigassocconttimecentindexfourth (\Yssconttimeval)}


\newcommand{\yssoutsepsystsigdisctimecentvec}
          {$\Yssoutsepsystsigdisctimecentvec$}
\newcommand{\Yssoutsepsystsigdisctimecentvec}
           {y}

\newcommand{\yssoutsepsystsigdisctimecentvecval}
          {$\Yssoutsepsystsigdisctimecentvecval$}
\newcommand{\Yssoutsepsystsigdisctimecentvecval}
           {\Yssoutsepsystsigdisctimecentvec (\Yssdisctimeval)}

\newcommand{\yssoutsepsystsigdisctimecentvecvalother}
{$\Yssoutsepsystsigdisctimecentvecvalother$}
\newcommand{\Yssoutsepsystsigdisctimecentvecvalother}
{\Yssoutsepsystsigdisctimecentvec (\Yssdisctimevalother)}

\newcommand{\yssoutsepsystsigdisctimecentone}
          {$\Yssoutsepsystsigdisctimecentone$}
\newcommand{\Yssoutsepsystsigdisctimecentone}
           {\Yssoutsepsystsigdisctimecentvec _{1}}

\newcommand{\yssoutsepsystsigdisctimecentoneval}
          {$\Yssoutsepsystsigdisctimecentoneval$}
\newcommand{\Yssoutsepsystsigdisctimecentoneval}
           {\Yssoutsepsystsigdisctimecentone (\Yssdisctimeval)}

\newcommand{\yssoutsepsystsigdisctimecentonevalother}
{$\Yssoutsepsystsigdisctimecentonevalother$}
\newcommand{\Yssoutsepsystsigdisctimecentonevalother}
{\Yssoutsepsystsigdisctimecentone (\Yssdisctimevalother)}

\newcommand{\yssoutsepsystsigdisctimecenttwo}
{$\Yssoutsepsystsigdisctimecenttwo$}
\newcommand{\Yssoutsepsystsigdisctimecenttwo}
{\Yssoutsepsystsigdisctimecentvec _{2}}

\newcommand{\yssoutsepsystsigdisctimecenttwovalother}
{$\Yssoutsepsystsigdisctimecenttwovalother$}
\newcommand{\Yssoutsepsystsigdisctimecenttwovalother}
{\Yssoutsepsystsigdisctimecenttwo (\Yssdisctimevalother)}

\newcommand{\yssoutsepsystsigdisctimecentindex}
          {$\Yssoutsepsystsigdisctimecentindex$}
\newcommand{\Yssoutsepsystsigdisctimecentindex}
           {\Yssoutsepsystsigdisctimecentvec _{\Ysswayindex}}

\newcommand{\yssoutsepsystsigdisctimecentindexval}
          {$\Yssoutsepsystsigdisctimecentindexval$}
\newcommand{\Yssoutsepsystsigdisctimecentindexval}
           {\Yssoutsepsystsigdisctimecentindex (\Yssdisctimeval)}

\newcommand{\yssoutsepsystsigdisctimecentindexvalother}
{$\Yssoutsepsystsigdisctimecentindexvalother$}
\newcommand{\Yssoutsepsystsigdisctimecentindexvalother}
{\Yssoutsepsystsigdisctimecentindex (\Yssdisctimevalother)}

\newcommand{\yssoutsepsystsigdisctimecentindexother}
          {$\Yssoutsepsystsigdisctimecentindexother$}
\newcommand{\Yssoutsepsystsigdisctimecentindexother}
           {\Yssoutsepsystsigdisctimecentvec _{\Ysswayindexother}}

\newcommand{\yssoutsepsystsigdisctimecentindexotherval}
          {$\Yssoutsepsystsigdisctimecentindexotherval$}
\newcommand{\Yssoutsepsystsigdisctimecentindexotherval}
           {\Yssoutsepsystsigdisctimecentindexother (\Yssdisctimeval)}

\newcommand{\yssoutsepsystsigdisctimecentindexothervalother}
{$\Yssoutsepsystsigdisctimecentindexothervalother$}
\newcommand{\Yssoutsepsystsigdisctimecentindexothervalother}
{\Yssoutsepsystsigdisctimecentindexother (\Yssdisctimevalother)}

\newcommand{\yssoutsepsystsigdisctimecentindexsensnb}
          {$\Yssoutsepsystsigdisctimecentindexsensnb$}
\newcommand{\Yssoutsepsystsigdisctimecentindexsensnb}
           {\Yssoutsepsystsigdisctimecentvec _{\Ysssensnb}}

\newcommand{\yssoutsepsystsigdisctimecentindexsensnbval}
          {$\Yssoutsepsystsigdisctimecentindexsensnbval$}
\newcommand{\Yssoutsepsystsigdisctimecentindexsensnbval}
           {\Yssoutsepsystsigdisctimecentindexsensnb (\Yssdisctimeval)}


\newcommand{\yssoutsepsystsigassocdisctimecontfreqnotonenoncentvec}
          {$\Yssoutsepsystsigassocdisctimecontfreqnotonenoncentvec$}
\newcommand{\Yssoutsepsystsigassocdisctimecontfreqnotonenoncentvec}
           {Y
       ^{\prime}}

\newcommand{\yssoutsepsystsigassocdisctimecontfreqnotonenoncentvecval}
          {$\Yssoutsepsystsigassocdisctimecontfreqnotonenoncentvecval$}
\newcommand{\Yssoutsepsystsigassocdisctimecontfreqnotonenoncentvecval}
           {\Yssoutsepsystsigassocdisctimecontfreqnotonenoncentvec ( \Ysscontfreqval ) }


\newcommand{\yssoutsepsystsigdisctimecentveczt}
          {$\Yssoutsepsystsigdisctimecentveczt$}
\newcommand{\Yssoutsepsystsigdisctimecentveczt}
           {Y}

\newcommand{\yssoutsepsystsigdisctimecentvecztval}
          {$\Yssoutsepsystsigdisctimecentvecztval$}
\newcommand{\Yssoutsepsystsigdisctimecentvecztval}
           {\Yssoutsepsystsigdisctimecentveczt (z)}

\newcommand{\yssoutsepsystsigdisctimecentonezt}
          {$\Yssoutsepsystsigdisctimecentonezt$}
\newcommand{\Yssoutsepsystsigdisctimecentonezt}
           {\Yssoutsepsystsigdisctimecentveczt _{1}}

\newcommand{\yssoutsepsystsigdisctimecentoneztval}
          {$\Yssoutsepsystsigdisctimecentoneztval$}
\newcommand{\Yssoutsepsystsigdisctimecentoneztval}
           {\Yssoutsepsystsigdisctimecentonezt (z)}

\newcommand{\yssoutsepsystsigdisctimecenttwozt}
          {$\Yssoutsepsystsigdisctimecenttwozt$}
\newcommand{\Yssoutsepsystsigdisctimecenttwozt}
           {\Yssoutsepsystsigdisctimecentveczt _{2}}

\newcommand{\yssoutsepsystsigdisctimecenttwoztval}
          {$\Yssoutsepsystsigdisctimecenttwoztval$}
\newcommand{\Yssoutsepsystsigdisctimecenttwoztval}
           {\Yssoutsepsystsigdisctimecenttwozt (z)}

\newcommand{\yssoutsepsystsigdisctimecentindexzt}
          {$\Yssoutsepsystsigdisctimecentindexzt$}
\newcommand{\Yssoutsepsystsigdisctimecentindexzt}
           {\Yssoutsepsystsigdisctimecentveczt _{\Ysswayindex}}

\newcommand{\yssoutsepsystsigdisctimecentindexztval}
          {$\Yssoutsepsystsigdisctimecentindexztval$}
\newcommand{\Yssoutsepsystsigdisctimecentindexztval}
           {\Yssoutsepsystsigdisctimecentindexzt (z)}

\newcommand{\yssoutsepsystsigdisctimecentindexsensnbzt}
          {$\Yssoutsepsystsigdisctimecentindexsensnbzt$}
\newcommand{\Yssoutsepsystsigdisctimecentindexsensnbzt}
           {\Yssoutsepsystsigdisctimecentveczt _{\Ysssensnb}}

\newcommand{\yssoutsepsystsigdisctimecentindexsensnbztval}
          {$\Yssoutsepsystsigdisctimecentindexsensnbztval$}
\newcommand{\Yssoutsepsystsigdisctimecentindexsensnbztval}
           {\Yssoutsepsystsigdisctimecentindexsensnbzt (z)}



\newcommand{\yssarbsigconttimenotone}
          {$\Yssarbsigconttimenotone$}
\newcommand{\Yssarbsigconttimenotone}
           {v}

\newcommand{\yssarbsigconttimenotoneval}
          {$\Yssarbsigconttimenotoneval$}
\newcommand{\Yssarbsigconttimenotoneval}
           {\Yssarbsigconttimenotone (\Yssconttimeval)}

\newcommand{\yssarbsigconttimenotoneindexone}
          {$\Yssarbsigconttimenotoneindexone$}
\newcommand{\Yssarbsigconttimenotoneindexone}
           {\Yssarbsigconttimenotone _{1}}

\newcommand{\yssarbsigconttimenotoneindexoneval}
          {$\Yssarbsigconttimenotoneindexoneval$}
\newcommand{\Yssarbsigconttimenotoneindexoneval}
           {\Yssarbsigconttimenotoneindexone (\Yssconttimeval)}

\newcommand{\yssarbsigconttimenotoneindextwo}
          {$\Yssarbsigconttimenotoneindextwo$}
\newcommand{\Yssarbsigconttimenotoneindextwo}
           {\Yssarbsigconttimenotone _{2}}

\newcommand{\yssarbsigconttimenotoneindextwoval}
          {$\Yssarbsigconttimenotoneindextwoval$}
\newcommand{\Yssarbsigconttimenotoneindextwoval}
           {\Yssarbsigconttimenotoneindextwo (\Yssconttimeval)}

\newcommand{\yssarbsigconttimenottwo}
          {$\Yssarbsigconttimenottwo$}
\newcommand{\Yssarbsigconttimenottwo}
           {w}


\newcommand{\yssarbsigconttimecontfreqnoncentnotone}
          {$\Yssarbsigconttimecontfreqnoncentnotone$}
\newcommand{\Yssarbsigconttimecontfreqnoncentnotone}
           {V}

\newcommand{\yssarbsigconttimecontfreqnoncentnotoneval}
          {$\Yssarbsigconttimecontfreqnoncentnotoneval$}
\newcommand{\Yssarbsigconttimecontfreqnoncentnotoneval}
           {\Yssarbsigconttimecontfreqnoncentnotone (\Yssconttimeval , \Ysscontfreqval)}

\newcommand{\yssarbsigconttimecontfreqnoncentnotoneindexone}
          {$\Yssarbsigconttimecontfreqnoncentnotoneindexone$}
\newcommand{\Yssarbsigconttimecontfreqnoncentnotoneindexone}
           {\Yssarbsigconttimecontfreqnoncentnotone _{1}}

\newcommand{\yssarbsigconttimecontfreqnoncentnotonevalindexone}
          {$\Yssarbsigconttimecontfreqnoncentnotonevalindexone$}
\newcommand{\Yssarbsigconttimecontfreqnoncentnotonevalindexone}
           {\Yssarbsigconttimecontfreqnoncentnotoneindexone (\Yssconttimeval , \Ysscontfreqval)}

\newcommand{\yssarbsigconttimecontfreqnoncentnotoneindextwo}
          {$\Yssarbsigconttimecontfreqnoncentnotoneindextwo$}
\newcommand{\Yssarbsigconttimecontfreqnoncentnotoneindextwo}
           {\Yssarbsigconttimecontfreqnoncentnotone _{2}}

\newcommand{\yssarbsigconttimecontfreqnoncentnotonevalindextwo}
          {$\Yssarbsigconttimecontfreqnoncentnotonevalindextwo$}
\newcommand{\Yssarbsigconttimecontfreqnoncentnotonevalindextwo}
           {\Yssarbsigconttimecontfreqnoncentnotoneindextwo (\Yssconttimeval , \Ysscontfreqval)}

\newcommand{\yssarbsigconttimecontfreqnoncentnottwo}
          {$\Yssarbsigconttimecontfreqnoncentnottwo$}
\newcommand{\Yssarbsigconttimecontfreqnoncentnottwo}
           {W}


\newcommand{\yssarbsigconttimecontscalenoncentnotone}
          {$\Yssarbsigconttimecontscalenoncentnotone$}
\newcommand{\Yssarbsigconttimecontscalenoncentnotone}
           {\Ysstimescalecontcoefnot _{\Yssarbsigconttimenotone}}

\newcommand{\yssarbsigconttimecontscalenoncentnotoneval}
          {$\Yssarbsigconttimecontscalenoncentnotoneval$}
\newcommand{\Yssarbsigconttimecontscalenoncentnotoneval}
           {\Yssarbsigconttimecontscalenoncentnotone
        ( \Ysstimescalecontshiftval , \Ysstimescalecontscaleval )
           }

\newcommand{\yssarbsigconttimecontscalenoncentnotoneindexone}
          {$\Yssarbsigconttimecontscalenoncentnotoneindexone$}
\newcommand{\Yssarbsigconttimecontscalenoncentnotoneindexone}
           {\Ysstimescalecontcoefnot _{\Yssarbsigconttimenotoneindexone} }

\newcommand{\yssarbsigconttimecontscalenoncentnotonevalindexone}
          {$\Yssarbsigconttimecontscalenoncentnotonevalindexone$}
\newcommand{\Yssarbsigconttimecontscalenoncentnotonevalindexone}
           {\Yssarbsigconttimecontscalenoncentnotoneindexone
        ( \Ysstimescalecontshiftval , \Ysstimescalecontscaleval )
       }

\newcommand{\yssarbsigconttimecontscalenoncentnotoneindextwo}
          {$\Yssarbsigconttimecontscalenoncentnotoneindextwo$}
\newcommand{\Yssarbsigconttimecontscalenoncentnotoneindextwo}
           {\Ysstimescalecontcoefnot _{\Yssarbsigconttimenotoneindextwo} }

\newcommand{\yssarbsigconttimecontscalenoncentnotonevalindextwo}
          {$\Yssarbsigconttimecontscalenoncentnotonevalindextwo$}
\newcommand{\Yssarbsigconttimecontscalenoncentnotonevalindextwo}
           {\Yssarbsigconttimecontscalenoncentnotoneindextwo
        ( \Ysstimescalecontshiftval , \Ysstimescalecontscaleval )
       }


\newcommand{\yssarbsigdisctimenotone}
          {$\Yssarbsigdisctimenotone$}
\newcommand{\Yssarbsigdisctimenotone}
           {v}

\newcommand{\yssarbsigdisctimenotoneval}
          {$\Yssarbsigdisctimenotoneval$}
\newcommand{\Yssarbsigdisctimenotoneval}
           {\Yssarbsigdisctimenotone (\Yssdisctimeval)}

\newcommand{\yssarbsigdisctimenottwo}
          {$\Yssarbsigdisctimenottwo$}
\newcommand{\Yssarbsigdisctimenottwo}
           {w}



\newcommand{\yssinnovtosrcfillengthnegnoindex}
           {$\Yssinnovtosrcfillengthnegnoindex$}
\newcommand{\Yssinnovtosrcfillengthnegnoindex}
           {L_{1}}

\newcommand{\yssinnovtosrcfillengthposnoindex}
           {$\Yssinnovtosrcfillengthposnoindex$}
\newcommand{\Yssinnovtosrcfillengthposnoindex}
           {L_{2}}


\newcommand{\yssinnovtosrcimprespnoindex}
           {$\Yssinnovtosrcimprespnoindex$}
\newcommand{\Yssinnovtosrcimprespnoindex}
           {d}

\newcommand{\yssinnovtosrcimprespone}
           {$\Yssinnovtosrcimprespone$}
\newcommand{\Yssinnovtosrcimprespone}
           {\Yssinnovtosrcimprespnoindex _{1}}

\newcommand{\yssinnovtosrcimpresptwo}
           {$\Yssinnovtosrcimpresptwo$}
\newcommand{\Yssinnovtosrcimpresptwo}
           {\Yssinnovtosrcimprespnoindex_{2}}

\newcommand{\yssinnovtosrcimprespindex}
           {$\Yssinnovtosrcimprespindex$}
\newcommand{\Yssinnovtosrcimprespindex}
           {\Yssinnovtosrcimprespnoindex_{\Ysswayindex}}


\newcommand{\yssinnovtosrctransfuncnoindex}
           {$\Yssinnovtosrctransfuncnoindex$}
\newcommand{\Yssinnovtosrctransfuncnoindex}
           {D}

\newcommand{\yssinnovtosrctransfuncindex}
           {$\Yssinnovtosrctransfuncindex$}
\newcommand{\Yssinnovtosrctransfuncindex}
           {\Yssinnovtosrctransfuncnoindex_{\Ysswayindex}}

\newcommand{\yssinnovtosrctransfuncindexval}
           {$\Yssinnovtosrctransfuncindexval$}
\newcommand{\Yssinnovtosrctransfuncindexval}
           {\Yssinnovtosrctransfuncindex (z)}



\newcommand{\yssmixmatrixscalar}
           {$\Yssmixmatrixscalar$}
\newcommand{\Yssmixmatrixscalar}
           {A}

\newcommand{\yssmixmatrixscalarestim}
           {$\Yssmixmatrixscalarestim$}
\newcommand{\Yssmixmatrixscalarestim}
           {\hat{\Yssmixmatrixscalar}}

\newcommand{\yssmixmatrixscalarelnoindex}
          {$\Yssmixmatrixscalarelnoindex$}
\newcommand{\Yssmixmatrixscalarelnoindex}
           {a}

\newcommand{\yssmixmatrixscalareloneone}
          {$\Yssmixmatrixscalareloneone$}
\newcommand{\Yssmixmatrixscalareloneone}
           {\Yssmixmatrixscalarelnoindex _{1 1}}

\newcommand{\yssmixmatrixscalarelonetwo}
          {$\Yssmixmatrixscalarelonetwo$}
\newcommand{\Yssmixmatrixscalarelonetwo}
           {\Yssmixmatrixscalarelnoindex _{1 2}}

\newcommand{\yssmixmatrixscalareloneindexother}
          {$\Yssmixmatrixscalareloneindexother$}
\newcommand{\Yssmixmatrixscalareloneindexother}
           {\Yssmixmatrixscalarelnoindex _{1 \Ysswayindexother}}

\newcommand{\yssmixmatrixscalareloneindexthird}
          {$\Yssmixmatrixscalareloneindexthird$}
\newcommand{\Yssmixmatrixscalareloneindexthird}
           {\Yssmixmatrixscalarelnoindex _{1 \Ysswayindexthird}}

\newcommand{\yssmixmatrixscalareloneindexfourth}
{$\Yssmixmatrixscalareloneindexfourth$}
\newcommand{\Yssmixmatrixscalareloneindexfourth}
{\Yssmixmatrixscalarelnoindex _{1 \Ysswayindexfourth}}

\newcommand{\yssmixmatrixscalareltwoone}
          {$\Yssmixmatrixscalareltwoone$}
\newcommand{\Yssmixmatrixscalareltwoone}
           {\Yssmixmatrixscalarelnoindex _{2 1}}

\newcommand{\yssmixmatrixscalareltwotwo}
          {$\Yssmixmatrixscalareltwotwo$}
\newcommand{\Yssmixmatrixscalareltwotwo}
           {\Yssmixmatrixscalarelnoindex _{2 2}}

\newcommand{\yssmixmatrixscalarelindexone}
          {$\Yssmixmatrixscalarelindexone$}
\newcommand{\Yssmixmatrixscalarelindexone}
           {\Yssmixmatrixscalarelnoindex _{\Ysswayindex 1}}

\newcommand{\yssmixmatrixscalarelindextwo}
          {$\Yssmixmatrixscalarelindextwo$}
\newcommand{\Yssmixmatrixscalarelindextwo}
           {\Yssmixmatrixscalarelnoindex _{\Ysswayindex 2}}

\newcommand{\yssmixmatrixscalarelindexthree}
          {$\Yssmixmatrixscalarelindexthree$}
\newcommand{\Yssmixmatrixscalarelindexthree}
           {\Yssmixmatrixscalarelnoindex _{\Ysswayindex 3}}

\newcommand{\yssmixmatrixscalarelindexindex}
          {$\Yssmixmatrixscalarelindexindex$}
\newcommand{\Yssmixmatrixscalarelindexindex}
           {\Yssmixmatrixscalarelnoindex _{\Ysswayindex \Ysswayindex}}

\newcommand{\yssmixmatrixscalarelindexindexother}
          {$\Yssmixmatrixscalarelindexindexother$}
\newcommand{\Yssmixmatrixscalarelindexindexother}
           {\Yssmixmatrixscalarelnoindex _{\Ysswayindex \Ysswayindexother}}

\newcommand{\yssmixmatrixscalarelindexindexthird}
          {$\Yssmixmatrixscalarelindexindexthird$}
\newcommand{\Yssmixmatrixscalarelindexindexthird}
           {\Yssmixmatrixscalarelnoindex _{\Ysswayindex \Ysswayindexthird}}

\newcommand{\yssmixmatrixscalarelindexindexfourth}
{$\Yssmixmatrixscalarelindexindexfourth$}
\newcommand{\Yssmixmatrixscalarelindexindexfourth}
{\Yssmixmatrixscalarelnoindex _{\Ysswayindex \Ysswayindexfourth}}

\newcommand{\yssmixmatrixscalarelindexotherindexother}
{$\Yssmixmatrixscalarelindexotherindexother$}
\newcommand{\Yssmixmatrixscalarelindexotherindexother}
{\Yssmixmatrixscalarelnoindex _{\Ysswayindexother \Ysswayindexother}}

\newcommand{\yssmixmatrixscalarelindexthirdindexother}
          {$\Yssmixmatrixscalarelindexthirdindexother$}
\newcommand{\Yssmixmatrixscalarelindexthirdindexother}
           {\Yssmixmatrixscalarelnoindex _{\Ysswayindexthird \Ysswayindexother}}

\newcommand{\yssmixmatrixscalarelindexthirdindexfourth}
          {$\Yssmixmatrixscalarelindexthirdindexfourth$}
\newcommand{\Yssmixmatrixscalarelindexthirdindexfourth}
           {\Yssmixmatrixscalarelnoindex _{\Ysswayindexthird \Ysswayindexfourth}}

\newcommand{\yssinvmixmatrixscalar}
           {$\Yssinvmixmatrixscalar$}
\newcommand{\Yssinvmixmatrixscalar}
           {\Yssmixmatrixscalar ^{-1}}

\newcommand{\yssinvmixmatrixscalarestim}
           {$\Yssinvmixmatrixscalarestim$}
\newcommand{\Yssinvmixmatrixscalarestim}
           {\Yssmixmatrixscalarestim ^{-1}}



\newcommand{\yssmixmatrixlagellagnoindex}
          {$\Yssmixmatrixlagellagnoindex$}
\newcommand{\Yssmixmatrixlagellagnoindex}
           {n}

\newcommand{\yssmixmatrixlagellagindexindexother}
          {$\Yssmixmatrixlagellagindexindexother$}
\newcommand{\Yssmixmatrixlagellagindexindexother}
           {\Yssmixmatrixlagellagnoindex _{\Ysswayindex \Ysswayindexother}}

\newcommand{\yssmixmatrixlagellagoneindexthird}
          {$\Yssmixmatrixlagellagoneindexthird$}
\newcommand{\Yssmixmatrixlagellagoneindexthird}
           {\Yssmixmatrixlagellagnoindex _{1 \Ysswayindexthird}}

\newcommand{\yssmixmatrixlagellagindexindexthird}
          {$\Yssmixmatrixlagellagindexindexthird$}
\newcommand{\Yssmixmatrixlagellagindexindexthird}
           {\Yssmixmatrixlagellagnoindex _{\Ysswayindex \Ysswayindexthird}}



\newcommand{\yssmixmatrixcontfreqnotonefunc}
           {$\Yssmixmatrixcontfreqnotonefunc$}
\newcommand{\Yssmixmatrixcontfreqnotonefunc}
           {A}


\newcommand{\yssmixmatrixcontfreqnotoneval}
           {$\Yssmixmatrixcontfreqnotoneval$}
\newcommand{\Yssmixmatrixcontfreqnotoneval}
           {\Yssmixmatrixcontfreqnotonefunc ( \Ysscontfreqval ) }




\newcommand{\yssmixtransfuncorder}
          {$\Yssmixtransfuncorder$}
\newcommand{\Yssmixtransfuncorder}
           {M}


\newcommand{\ymixmatrixzt}
           {$\Ymixmatrixzt$}
\newcommand{\Ymixmatrixzt}
           {A(z)}


\newcommand{\yssmixtransfuncnoindex}
           {$\Yssmixtransfuncnoindex$}
\newcommand{\Yssmixtransfuncnoindex}
           {A}

\newcommand{\yssmixtransfunconeone}
          {$\Yssmixtransfunconeone$}
\newcommand{\Yssmixtransfunconeone}
           {\Yssmixtransfuncnoindex _{1 1} (z)}

\newcommand{\yssmixtransfunconetwo}
          {$\Yssmixtransfunconetwo$}
\newcommand{\Yssmixtransfunconetwo}
           {\Yssmixtransfuncnoindex _{1 2} (z)}

\newcommand{\yssmixtransfunconethree}
          {$\Yssmixtransfunconethree$}
\newcommand{\Yssmixtransfunconethree}
           {\Yssmixtransfuncnoindex _{1 3} (z)}

\newcommand{\yssmixtransfunctwoone}
          {$\Yssmixtransfunctwoone$}
\newcommand{\Yssmixtransfunctwoone}
           {\Yssmixtransfuncnoindex _{2 1} (z)}

\newcommand{\yssmixtransfunctwotwo}
          {$\Yssmixtransfunctwotwo$}
\newcommand{\Yssmixtransfunctwotwo}
           {\Yssmixtransfuncnoindex _{2 2} (z)}

\newcommand{\yssmixtransfunctwothree}
          {$\Yssmixtransfunctwothree$}
\newcommand{\Yssmixtransfunctwothree}
           {\Yssmixtransfuncnoindex _{2 3} (z)}

\newcommand{\yssmixtransfunconeindexother}
          {$\Yssmixtransfunconeindexother$}
\newcommand{\Yssmixtransfunconeindexother}
           {\Yssmixtransfuncnoindex _{1 \Ysswayindexother} (z)}

\newcommand{\yssmixtransfunctwoindexother}
          {$\Yssmixtransfunctwoindexother$}
\newcommand{\Yssmixtransfunctwoindexother}
           {\Yssmixtransfuncnoindex _{2 \Ysswayindexother} (z)}

\newcommand{\yssmixtransfuncindexindexother}
          {$\Yssmixtransfuncindexindexother$}
\newcommand{\Yssmixtransfuncindexindexother}
           {\Yssmixtransfuncnoindex _{\Ysswayindex \Ysswayindexother} (z)}


\newcommand{\yssmixmatrixscalarquad}
{$\Yssmixmatrixscalarquad$}
\newcommand{\Yssmixmatrixscalarquad}
{Q}

\newcommand{\yssmixmatrixscalarelquadnoindex}
          {$\Yssmixmatrixscalarelquadnoindex$}
\newcommand{\Yssmixmatrixscalarelquadnoindex}
           {q}

\newcommand{\yssmixmatrixscalarelquadone}
{$\Yssmixmatrixscalarelquadone$}
\newcommand{\Yssmixmatrixscalarelquadone}
{\Yssmixmatrixscalarelquadnoindex _{1}}

\newcommand{\yssmixmatrixscalarelquadtwo}
{$\Yssmixmatrixscalarelquadtwo$}
\newcommand{\Yssmixmatrixscalarelquadtwo}
{\Yssmixmatrixscalarelquadnoindex _{2}}

\newcommand{\yssmixmatrixscalarelquadindex}
{$\Yssmixmatrixscalarelquadindex$}
\newcommand{\Yssmixmatrixscalarelquadindex}
{\Yssmixmatrixscalarelquadnoindex _{\Ysswayindex}}

\newcommand{\yssmixmatrixscalarelquadoneindexotherindexthird}
{$\Yssmixmatrixscalarelquadoneindexotherindexthird$}
\newcommand{\Yssmixmatrixscalarelquadoneindexotherindexthird}
{\Yssmixmatrixscalarelquadnoindex _{1 \Ysswayindexother \Ysswayindexthird}}

\newcommand{\yssmixmatrixscalarelquadindexindexotherindexthird}
{$\Yssmixmatrixscalarelquadindexindexotherindexthird$}
\newcommand{\Yssmixmatrixscalarelquadindexindexotherindexthird}
{\Yssmixmatrixscalarelquadnoindex _{\Ysswayindex \Ysswayindexother \Ysswayindexthird}}


\newcommand{\yssmixmatrixscalarlinquadpartlinelnoindex}
{$\Yssmixmatrixscalarlinquadpartlinelnoindex$}
\newcommand{\Yssmixmatrixscalarlinquadpartlinelnoindex}
{L}

\newcommand{\yssmixmatrixscalarlinquadpartlinelonetwo}
{$\Yssmixmatrixscalarlinquadpartlinelonetwo$}
\newcommand{\Yssmixmatrixscalarlinquadpartlinelonetwo}
{\Yssmixmatrixscalarlinquadpartlinelnoindex_{12}}

\newcommand{\yssmixmatrixscalarlinquadpartlineltwoone}
{$\Yssmixmatrixscalarlinquadpartlineltwoone$}
\newcommand{\Yssmixmatrixscalarlinquadpartlineltwoone}
{\Yssmixmatrixscalarlinquadpartlinelnoindex_{21}}

\newcommand{\yssmixmatrixscalarlinquadpartlinelindexindexother}
{$\Yssmixmatrixscalarlinquadpartlinelindexindexother$}
\newcommand{\Yssmixmatrixscalarlinquadpartlinelindexindexother}
{\Yssmixmatrixscalarlinquadpartlinelnoindex_{\Ysswayindex \Ysswayindexother}}

\newcommand{\yssmixmatrixscalarlinquadpartquadelnoindex}
{$\Yssmixmatrixscalarlinquadpartquadelnoindex$}
\newcommand{\Yssmixmatrixscalarlinquadpartquadelnoindex}
{Q}

\newcommand{\yssmixmatrixscalarlinquadpartquadelone}
{$\Yssmixmatrixscalarlinquadpartquadelone$}
\newcommand{\Yssmixmatrixscalarlinquadpartquadelone}
{\Yssmixmatrixscalarlinquadpartquadelnoindex_{1}}

\newcommand{\yssmixmatrixscalarlinquadpartquadeltwo}
{$\Yssmixmatrixscalarlinquadpartquadeltwo$}
\newcommand{\Yssmixmatrixscalarlinquadpartquadeltwo}
{\Yssmixmatrixscalarlinquadpartquadelnoindex_{2}}

\newcommand{\yssmixmatrixscalarlinquadpartquadelindex}
{$\Yssmixmatrixscalarlinquadpartquadelindex$}
\newcommand{\Yssmixmatrixscalarlinquadpartquadelindex}
{\Yssmixmatrixscalarlinquadpartquadelnoindex_{\Ysswayindex}}



\newcommand{\ysssepsystmatrixscalar}
          {$\Ysssepsystmatrixscalar$}
\newcommand{\Ysssepsystmatrixscalar}
           {B}

\newcommand{\ysssepsystmatrixscalarelnoindex}
          {$\Ysssepsystmatrixscalarelnoindex$}
\newcommand{\Ysssepsystmatrixscalarelnoindex}
           {b}

\newcommand{\ysssepsystmatrixscalareloneindexother}
          {$\Ysssepsystmatrixscalareloneindexother$}
\newcommand{\Ysssepsystmatrixscalareloneindexother}
           {\Ysssepsystmatrixscalarelnoindex _{1 \Ysswayindexother}}

\newcommand{\ysssepsystmatrixscalarelindexindexother}
          {$\Ysssepsystmatrixscalarelindexindexother$}
\newcommand{\Ysssepsystmatrixscalarelindexindexother}
           {\Ysssepsystmatrixscalarelnoindex _{\Ysswayindex \Ysswayindexother}}


\newcommand{\ysssepsysmatrixscalarassocone}
           {$\Ysssepsysmatrixscalarassocone$}
\newcommand{\Ysssepsysmatrixscalarassocone}
           {C}

\newcommand{\ysssepsystmatrixscalarassoconeelnoindex}
          {$\Ysssepsystmatrixscalarassoconeelnoindex$}
\newcommand{\Ysssepsystmatrixscalarassoconeelnoindex}
           {\alpha}

\newcommand{\ysssepsystmatrixscalarassoconeeltwoone}
          {$\Ysssepsystmatrixscalarassoconeeltwoone$}
\newcommand{\Ysssepsystmatrixscalarassoconeeltwoone}
           {\Ysssepsystmatrixscalarassoconeelnoindex _{2 1}}

\newcommand{\ysssepsystmatrixscalarassoconeeltwotwo}
          {$\Ysssepsystmatrixscalarassoconeeltwotwo$}
\newcommand{\Ysssepsystmatrixscalarassoconeeltwotwo}
           {\Ysssepsystmatrixscalarassoconeelnoindex _{2 2}}

\newcommand{\ysssepsystmatrixscalarassoconeelindexindexother}
          {$\Ysssepsystmatrixscalarassoconeelindexindexother$}
\newcommand{\Ysssepsystmatrixscalarassoconeelindexindexother}
           {\Ysssepsystmatrixscalarassoconeelnoindex _{\Ysswayindex \Ysswayindexother}}

\newcommand{\ysssepsystmatrixscalarassoconeelindexindexthird}
          {$\Ysssepsystmatrixscalarassoconeelindexindexthird$}
\newcommand{\Ysssepsystmatrixscalarassoconeelindexindexthird}
           {\Ysssepsystmatrixscalarassoconeelnoindex _{\Ysswayindex \Ysswayindexthird}}


\newcommand{\ysssepsysmatrixscalarassoctwo}
           {$\Ysssepsysmatrixscalarassoctwo$}
\newcommand{\Ysssepsysmatrixscalarassoctwo}
           {D}

\newcommand{\ysssepsystmatrixscalarassoctwoelnoindex}
          {$\Ysssepsystmatrixscalarassoctwoelnoindex$}
\newcommand{\Ysssepsystmatrixscalarassoctwoelnoindex}
           {\beta}

\newcommand{\ysssepsystmatrixscalarassoctwoeltwoone}
          {$\Ysssepsystmatrixscalarassoctwoeltwoone$}
\newcommand{\Ysssepsystmatrixscalarassoctwoeltwoone}
           {\Ysssepsystmatrixscalarassoctwoelnoindex _{2 1}}

\newcommand{\ysssepsystmatrixscalarassoctwoeltwotwo}
          {$\Ysssepsystmatrixscalarassoctwoeltwotwo$}
\newcommand{\Ysssepsystmatrixscalarassoctwoeltwotwo}
           {\Ysssepsystmatrixscalarassoctwoelnoindex _{2 2}}

\newcommand{\ysssepsystmatrixscalarassoctwoelindexindexother}
          {$\Ysssepsystmatrixscalarassoctwoelindexindexother$}
\newcommand{\Ysssepsystmatrixscalarassoctwoelindexindexother}
           {\Ysssepsystmatrixscalarassoctwoelnoindex _{\Ysswayindex \Ysswayindexother}}

\newcommand{\ysssepsystmatrixscalarassoctwoelindexindexthird}
          {$\Ysssepsystmatrixscalarassoctwoelindexindexthird$}
\newcommand{\Ysssepsystmatrixscalarassoctwoelindexindexthird}
           {\Ysssepsystmatrixscalarassoctwoelnoindex _{\Ysswayindex \Ysswayindexthird}}


\newcommand{\ysssepsysmatrixscalarassocthree}
           {$\Ysssepsysmatrixscalarassocthree$}
\newcommand{\Ysssepsysmatrixscalarassocthree}
           {\Ysssepsysmatrixscalarassocone^{\prime}}

\newcommand{\ysssepsystmatrixscalarassocthreeelnoindex}
          {$\Ysssepsystmatrixscalarassocthreeelnoindex$}
\newcommand{\Ysssepsystmatrixscalarassocthreeelnoindex}
           {\Ysssepsystmatrixscalarassoconeelnoindex^{\prime}}

\newcommand{\ysssepsystmatrixscalarassocthreeelindexindexother}
          {$\Ysssepsystmatrixscalarassocthreeelindexindexother$}
\newcommand{\Ysssepsystmatrixscalarassocthreeelindexindexother}
           {\Ysssepsystmatrixscalarassocthreeelnoindex _{\Ysswayindex \Ysswayindexother}}

\newcommand{\ysssepsystmatrixscalarassocthreeelindexindexthird}
          {$\Ysssepsystmatrixscalarassocthreeelindexindexthird$}
\newcommand{\Ysssepsystmatrixscalarassocthreeelindexindexthird}
           {\Ysssepsystmatrixscalarassocthreeelnoindex _{\Ysswayindex \Ysswayindexthird}}


\newcommand{\ysssepsystmatrixcontfreqnotonefunc}
           {$\Ysssepsystmatrixcontfreqnotonefunc$}
\newcommand{\Ysssepsystmatrixcontfreqnotonefunc}
           {B}

\newcommand{\ysssepsystmatrixcontfreqnotoneval}
           {$\Ysssepsystmatrixcontfreqnotoneval$}
\newcommand{\Ysssepsystmatrixcontfreqnotoneval}
           {\Ysssepsystmatrixcontfreqnotonefunc ( \Ysscontfreqval ) }

\newcommand{\ysssepsystindexindexothercontfreqnotoneval}
          {$\Ysssepsystindexindexothercontfreqnotoneval$}
\newcommand{\Ysssepsystindexindexothercontfreqnotoneval}
           {\Ysssepsystmatrixcontfreqnotonefunc
        _{\Ysswayindex \Ysswayindexother} ( \Ysscontfreqval )}




\newcommand{\ysssepsysttransfuncnoindex}
           {$\Ysssepsysttransfuncnoindex$}
\newcommand{\Ysssepsysttransfuncnoindex}
           {B}

\newcommand{\ysssepsysttransfuncnoindexval}
           {$\Ysssepsysttransfuncnoindexval$}
\newcommand{\Ysssepsysttransfuncnoindexval}
           {\Ysssepsysttransfuncnoindex (z)}

\newcommand{\ysssepsysttransfunconetwo}
          {$\Ysssepsysttransfunconetwo$}
\newcommand{\Ysssepsysttransfunconetwo}
           {\Ysssepsysttransfuncnoindex _{1 2} (z)}

\newcommand{\ysssepsysttransfunctwoone}
          {$\Ysssepsysttransfunctwoone$}
\newcommand{\Ysssepsysttransfunctwoone}
           {\Ysssepsysttransfuncnoindex _{2 1} (z)}

\newcommand{\ysssepsysttransfuncindexindexother}
          {$\Ysssepsysttransfuncindexindexother$}
\newcommand{\Ysssepsysttransfuncindexindexother}
           {\Ysssepsysttransfuncnoindex _{\Ysswayindex \Ysswayindexother} (z)}


\newcommand{\ysssepsystcoefnoindex}
           {$\Ysssepsystcoefnoindex$}
\newcommand{\Ysssepsystcoefnoindex}
           {b}

\newcommand{\ysssepsystcoefindexindexother}
           {$\Ysssepsystcoefindexindexother$}
\newcommand{\Ysssepsystcoefindexindexother}
           {\Ysssepsystcoefnoindex _{\Ysswayindex \Ysswayindexother}}

\newcommand{\ysssepsystcoefonetwo}
           {$\Ysssepsystcoefonetwo$}
\newcommand{\Ysssepsystcoefonetwo}
           {\Ysssepsystcoefnoindex _{1 2}}

\newcommand{\ysssepsystcoeftwoone}
           {$\Ysssepsystcoeftwoone$}
\newcommand{\Ysssepsystcoeftwoone}
           {\Ysssepsystcoefnoindex _{2 1}}


\newcommand{\ysssepsystscalarlinquadpartlinelnoindex}
{$\Ysssepsystscalarlinquadpartlinelnoindex$}
\newcommand{\Ysssepsystscalarlinquadpartlinelnoindex}
{l}

\newcommand{\ysssepsystscalarlinquadpartlineloneone}
{$\Ysssepsystscalarlinquadpartlineloneone$}
\newcommand{\Ysssepsystscalarlinquadpartlineloneone}
{\Ysssepsystscalarlinquadpartlinelnoindex_{11}}

\newcommand{\ysssepsystscalarlinquadpartlineloneoneassoc}
{$\Ysssepsystscalarlinquadpartlineloneoneassoc$}
\newcommand{\Ysssepsystscalarlinquadpartlineloneoneassoc}
{\Ysssepsystscalarlinquadpartlineloneone ^{\prime}}

\newcommand{\ysssepsystscalarlinquadpartlinelonetwo}
{$\Ysssepsystscalarlinquadpartlinelonetwo$}
\newcommand{\Ysssepsystscalarlinquadpartlinelonetwo}
{\Ysssepsystscalarlinquadpartlinelnoindex_{12}}

\newcommand{\ysssepsystscalarlinquadpartlineltwoone}
{$\Ysssepsystscalarlinquadpartlineltwoone$}
\newcommand{\Ysssepsystscalarlinquadpartlineltwoone}
{\Ysssepsystscalarlinquadpartlinelnoindex_{21}}

\newcommand{\ysssepsystscalarlinquadpartlineltwotwo}
{$\Ysssepsystscalarlinquadpartlineltwotwo$}
\newcommand{\Ysssepsystscalarlinquadpartlineltwotwo}
{\Ysssepsystscalarlinquadpartlinelnoindex_{22}}

\newcommand{\ysssepsystscalarlinquadpartlineltwotwoassoc}
{$\Ysssepsystscalarlinquadpartlineltwotwoassoc$}
\newcommand{\Ysssepsystscalarlinquadpartlineltwotwoassoc}
{\Ysssepsystscalarlinquadpartlineltwotwo ^{\prime}}

\newcommand{\ysssepsystscalarlinquadpartlinelindexindex}
{$\Ysssepsystscalarlinquadpartlinelindexindex$}
\newcommand{\Ysssepsystscalarlinquadpartlinelindexindex}
{\Ysssepsystscalarlinquadpartlinelnoindex_{\Ysswayindex \Ysswayindex}}

\newcommand{\ysssepsystscalarlinquadpartlinelindexindexassoc}
{$\Ysssepsystscalarlinquadpartlinelindexindexassoc$}
\newcommand{\Ysssepsystscalarlinquadpartlinelindexindexassoc}
{\Ysssepsystscalarlinquadpartlinelindexindex ^{\prime}}

\newcommand{\ysssepsystscalarlinquadpartlinelindexindexother}
{$\Ysssepsystscalarlinquadpartlinelindexindexother$}
\newcommand{\Ysssepsystscalarlinquadpartlinelindexindexother}
{\Ysssepsystscalarlinquadpartlinelnoindex_{\Ysswayindex \Ysswayindexother}}

\newcommand{\ysssepsystscalarlinquadpartquadelnoindex}
{$\Ysssepsystscalarlinquadpartquadelnoindex$}
\newcommand{\Ysssepsystscalarlinquadpartquadelnoindex}
{q}

\newcommand{\ysssepsystscalarlinquadpartquadelone}
{$\Ysssepsystscalarlinquadpartquadelone$}
\newcommand{\Ysssepsystscalarlinquadpartquadelone}
{\Ysssepsystscalarlinquadpartquadelnoindex_{1}}

\newcommand{\ysssepsystscalarlinquadpartquadeltwo}
{$\Ysssepsystscalarlinquadpartquadeltwo$}
\newcommand{\Ysssepsystscalarlinquadpartquadeltwo}
{\Ysssepsystscalarlinquadpartquadelnoindex_{2}}

\newcommand{\ysssepsystscalarlinquadpartquadelindex}
{$\Ysssepsystscalarlinquadpartquadelindex$}
\newcommand{\Ysssepsystscalarlinquadpartquadelindex}
{\Ysssepsystscalarlinquadpartquadelnoindex_{\Ysswayindex}}



\newcommand{\ysscomplfiltcoefnoindex}
           {$\Ysscomplfiltcoefnoindex$}
\newcommand{\Ysscomplfiltcoefnoindex}
           {h}

\newcommand{\ysscomplfiltcoefindexone}
           {$\Ysscomplfiltcoefindexone$}
\newcommand{\Ysscomplfiltcoefindexone}
           {\Ysscomplfiltcoefnoindex _{1}}

\newcommand{\ysscomplfiltcoefindextwo}
           {$\Ysscomplfiltcoefindextwo$}
\newcommand{\Ysscomplfiltcoefindextwo}
           {\Ysscomplfiltcoefnoindex _{2}}

\newcommand{\ysscomplfiltcoefindexindex}
           {$\Ysscomplfiltcoefindexindex$}
\newcommand{\Ysscomplfiltcoefindexindex}
           {\Ysscomplfiltcoefnoindex _{\Ysswayindex}}

\newcommand{\ysscomplfiltcoefindexindexthird}
           {$\Ysscomplfiltcoefindexindexthird$}
\newcommand{\Ysscomplfiltcoefindexindexthird}
           {\Ysscomplfiltcoefnoindex _{\Ysswayindex \Ysswayindexthird}}

\newcommand{\ysscomplfiltcoefindexotherindexthird}
           {$\Ysscomplfiltcoefindexotherindexthird$}
\newcommand{\Ysscomplfiltcoefindexotherindexthird}
           {\Ysscomplfiltcoefnoindex _{\Ysswayindexother \Ysswayindexthird}}



\newcommand{\ysssepsystadaptgainnoindex}
           {$\Ysssepsystadaptgainnoindex$}
\newcommand{\Ysssepsystadaptgainnoindex}
           {\mu}

\newcommand{\ysssepsystadaptgainone}
           {$\Ysssepsystadaptgainone$}
\newcommand{\Ysssepsystadaptgainone}
           {\Ysssepsystadaptgainnoindex _{1}}

\newcommand{\ysssepsystadaptgaintwo}
           {$\Ysssepsystadaptgaintwo$}
\newcommand{\Ysssepsystadaptgaintwo}
           {\Ysssepsystadaptgainnoindex _{2}}

\newcommand{\ysssepsystadaptgainindex}
           {$\Ysssepsystadaptgainindex$}
\newcommand{\Ysssepsystadaptgainindex}
           {\Ysssepsystadaptgainnoindex _{\Ysswayindex}}



\newcommand{\yodedisctimeval}
          {$\Yodedisctimeval$}
\newcommand{\Yodedisctimeval}
           {\Yssdisctimeval}


\newcommand{\yodesystemstatenoindex}
          {$\Yodesystemstatenoindex$}
\newcommand{\Yodesystemstatenoindex}
           {\theta}

\newcommand{\yodesystemstateindex}
          {$\Yodesystemstateindex$}
\newcommand{\Yodesystemstateindex}
           {\Yodesystemstatenoindex _{\Yodedisctimeval}}

\newcommand{\yodesystemstateeq}
          {$\Yodesystemstateeq$}
\newcommand{\Yodesystemstateeq}
           {\Yodesystemstatenoindex ^{*}}

\newcommand{\yodesystemstatesep}
          {$\Yodesystemstatesep$}
\newcommand{\Yodesystemstatesep}
           {\Yodesystemstatenoindex ^{s}}


\newcommand{\yodesystemsignoindex}
          {$\Yodesystemsignoindex$}
\newcommand{\Yodesystemsignoindex}
           {\xi}

\newcommand{\yodesystemsigtimeindexplusone}
          {$\Yodesystemsigtimeindexplusone$}
\newcommand{\Yodesystemsigtimeindexplusone}
           {\Yodesystemsignoindex _{\Yodedisctimeval + 1}}


\newcommand{\yodesystemfuncnoindex}
          {$\Yodesystemfuncnoindex$}
\newcommand{\Yodesystemfuncnoindex}
           {H}

\newcommand{\yodesystemfuncindex}
          {$\Yodesystemfuncindex$}
\newcommand{\Yodesystemfuncindex}
           {\Yodesystemfuncnoindex ( \Yodesystemstateindex , \Yodesystemsigtimeindexplusone) }


\newcommand{\yodesystemjacobmatrix}
          {$\Yodesystemjacobmatrix$}
\newcommand{\Yodesystemjacobmatrix}
           {J}

\newcommand{\yodesystemjacobmatrixstatesep}
          {$\Yodesystemjacobmatrixstatesep$}
\newcommand{\Yodesystemjacobmatrixstatesep}
           {\Yodesystemjacobmatrix ( \Yodesystemstatesep )}

\newcommand{\yodesystemjacobmatrixlineindex}
          {$\Yodesystemjacobmatrixlineindex$}
\newcommand{\Yodesystemjacobmatrixlineindex}
           {i}

\newcommand{\yodesystemjacobmatrixcolumnindex}
          {$\Yodesystemjacobmatrixcolumnindex$}
\newcommand{\Yodesystemjacobmatrixcolumnindex}
           {j}

\newcommand{\yodesystemjacobmatrixellineindexcolumnindex}
          {$\Yodesystemjacobmatrixellineindexcolumnindex$}
\newcommand{\Yodesystemjacobmatrixellineindexcolumnindex}
           {\Yodesystemjacobmatrix _{\Yodesystemjacobmatrixlineindex \Yodesystemjacobmatrixcolumnindex}}

%% file: artitd47_rapport_v4_com.tex
%
%
%
%
%
%
\newcommand{\ywritereadonestatenb}
{$\Ywritereadonestatenb$}
\newcommand{\Ywritereadonestatenb}
{K}
%
\newcommand{\ywritereadonestatestepindex}
{$\Ywritereadonestatestepindex$}
\newcommand{\Ywritereadonestatestepindex}
{k}
%
%
%
%
\newcommand{\ytwoqubitseqindex}
{$\Ytwoqubitseqindex$}
\newcommand{\Ytwoqubitseqindex}
{n}
%
%
%
%
%
\newcommand{\yonequbittimewrite}
{$\Yonequbittimewrite$}
\newcommand{\Yonequbittimewrite}
{t_w}
%
\newcommand{\yonequbittimeread}
{$\Yonequbittimeread$}
\newcommand{\Yonequbittimeread}
{t_r}
%
%
%
%
\newcommand{\yonequbittimewritestepindex}
{$\Yonequbittimewritestepindex$}
\newcommand{\Yonequbittimewritestepindex}
{\Yonequbittimewrite (\Ywritereadonestatestepindex)}
%
\newcommand{\yonequbittimereadstepindex}
{$\Yonequbittimereadstepindex$}
\newcommand{\Yonequbittimereadstepindex}
{\Yonequbittimeread (\Ywritereadonestatestepindex)}
%
%
\newcommand{\ytwoqubitwritereadtimeinterval}
{$\Ytwoqubitwritereadtimeinterval$}
\newcommand{\Ytwoqubitwritereadtimeinterval}
{T}
%
\newcommand{\ytwoqubitwritereadtimeintervalindex}
{$\Ytwoqubitwritereadtimeintervalindex$}
\newcommand{\Ytwoqubitwritereadtimeintervalindex}
{\Ytwoqubitwritereadtimeinterval (\Ywritereadonestatestepindex)}
%
%
%
%
\newcommand{\ytwoqubitsbasisplusplus}
{$\Ytwoqubitsbasisplusplus$}
\newcommand{\Ytwoqubitsbasisplusplus}
{{\cal B} _+}
%
\newcommand{\ytwoqubitsbasisoneone}
{$\Ytwoqubitsbasisoneone$}
\newcommand{\Ytwoqubitsbasisoneone}
{{\cal B} _1}
%
%
%
%
\newcommand{\ytwoqubitsprobaplusplus}
{$\Ytwoqubitsprobaplusplus$}
\newcommand{\Ytwoqubitsprobaplusplus}
{p_1}
%
\newcommand{\ytwoqubitsprobaminusminus}
{$\Ytwoqubitsprobaminusminus$}
\newcommand{\Ytwoqubitsprobaminusminus}
{p_2}
%
\newcommand{\ytwoqubitsprobaplusminus}
{$\Ytwoqubitsprobaplusminus$}
\newcommand{\Ytwoqubitsprobaplusminus}
{p_3}
%
\newcommand{\ytwoqubitsprobaminusplus}
{$\Ytwoqubitsprobaminusplus$}
\newcommand{\Ytwoqubitsprobaminusplus}
{p_4}
%
\newcommand{\ytwoqubitsprobaindexstd}
{$\Ytwoqubitsprobaindexstd$}
\newcommand{\Ytwoqubitsprobaindexstd}
{p_j}

%
%
%
%
\newcommand{\yqubitindexstd}
{$\Yqubitindexstd$}
\newcommand{\Yqubitindexstd}
{i}
%
%
%
%
\newcommand{\yparamqubitbothstateplusmodulusnot}
{$\Yparamqubitbothstateplusmodulusnot$}
\newcommand{\Yparamqubitbothstateplusmodulusnot}
{r}
%
\newcommand{\yparamqubitonestateplusmodulus}
{$\Yparamqubitonestateplusmodulus$}
\newcommand{\Yparamqubitonestateplusmodulus}
{{\Yparamqubitbothstateplusmodulusnot}_1}
%
\newcommand{\yparamqubittwostateplusmodulus}
{$\Yparamqubittwostateplusmodulus$}
\newcommand{\Yparamqubittwostateplusmodulus}
{{\Yparamqubitbothstateplusmodulusnot}_2}
%
\newcommand{\yparamqubitindexstdstateplusmodulus}
{$\Yparamqubitindexstdstateplusmodulus$}
\newcommand{\Yparamqubitindexstdstateplusmodulus}
{{\Yparamqubitbothstateplusmodulusnot}_{\Yqubitindexstd}}
%
\newcommand{\yparamqubitbothstateminusmodulusnot}
{$\Yparamqubitbothstateminusmodulusnot$}
\newcommand{\Yparamqubitbothstateminusmodulusnot}
{q}
%
\newcommand{\yparamqubitonestateminusmodulus}
{$\Yparamqubitonestateminusmodulus$}
\newcommand{\Yparamqubitonestateminusmodulus}
{{\Yparamqubitbothstateminusmodulusnot}_1}
%
\newcommand{\yparamqubittwostateminusmodulus}
{$\Yparamqubittwostateminusmodulus$}
\newcommand{\Yparamqubittwostateminusmodulus}
{{\Yparamqubitbothstateminusmodulusnot}_2}
%
\newcommand{\yparamqubitindexstdstateminusmodulus}
{$\Yparamqubitindexstdstateminusmodulus$}
\newcommand{\Yparamqubitindexstdstateminusmodulus}
{{\Yparamqubitbothstateminusmodulusnot}_{\Yqubitindexstd}}
%
\newcommand{\yparamqubitonestateplusphase}
{$\Yparamqubitonestateplusphase$}
\newcommand{\Yparamqubitonestateplusphase}
{\theta_1}
%
\newcommand{\yparamqubittwostateplusphase}
{$\Yparamqubittwostateplusphase$}
\newcommand{\Yparamqubittwostateplusphase}
{\theta_2}
%
\newcommand{\yparamqubitonestateminusphase}
{$\Yparamqubitonestateminusphase$}
\newcommand{\Yparamqubitonestateminusphase}
{\phi_1}
%
\newcommand{\yparamqubittwostateminusphase}
{$\Yparamqubittwostateminusphase$}
\newcommand{\Yparamqubittwostateminusphase}
{\phi_2}
%
%
%
%
\newcommand{\ytwoqubitresultphaseinit}
{$\Ytwoqubitresultphaseinit$}
\newcommand{\Ytwoqubitresultphaseinit}
{\Delta _I}
%
\newcommand{\ytwoqubitresultphaseevol}
{$\Ytwoqubitresultphaseevol$}
\newcommand{\Ytwoqubitresultphaseevol}
{\Delta _E}
%
\newcommand{\ytwoqubitresultphaseevolsin}
{$\Ytwoqubitresultphaseevolsin$}
\newcommand{\Ytwoqubitresultphaseevolsin}
{v}
%
%
%
%
\newcommand{\ymagfieldnot}
{$\Ymagfieldnot$}
\newcommand{\Ymagfieldnot}
{B}
%
%
%
\newcommand{\yhamiltonfieldscale}
{$\Yhamiltonfieldscale$}
\newcommand{\Yhamiltonfieldscale}
{G}
%
%
%
%
\newcommand{\ymixfunc}
{$\Ymixfunc$}
\newcommand{\Ymixfunc}
{g}
%
%
%
%
\newcommand{\ymixfuncimpldiff}
{$\Ymixfuncimpldiff$}
\newcommand{\Ymixfuncimpldiff}
{F}
%
%
%
\newcommand{\ymixfuncjacob}
{$\Ymixfuncjacob$}
\newcommand{\Ymixfuncjacob}
{J_{\Ymixfunc}}
%
\newcommand{\ymixfuncjacobval}
{$\Ymixfuncjacobval$}
\newcommand{\Ymixfuncjacobval}
{\Ymixfuncjacob ( \Ysssrcsigdisctimecentvec )}
%
%
%
%
\newcommand{\yparamqubitindexstdstateplusmodulussign}
{$\Yparamqubitindexstdstateplusmodulussign$}
\newcommand{\Yparamqubitindexstdstateplusmodulussign}
{\epsilon}
%
%
%
%
%
\newcommand{\ysssrcsigdisctimecentvecrand}
{$\Ysssrcsigdisctimecentvecrand$}
\newcommand{\Ysssrcsigdisctimecentvecrand}
{S}
%
\newcommand{\ysssrcsigdisctimecentindexrand}
{$\Ysssrcsigdisctimecentindexrand$}
\newcommand{\Ysssrcsigdisctimecentindexrand}
{\Ysssrcsigdisctimecentvecrand _ {\Ysswayindex}}
%
%
%
\newcommand{\ysssrcsigdisctimecentvecranddens}
{$\Ysssrcsigdisctimecentvecranddens$}
\newcommand{\Ysssrcsigdisctimecentvecranddens}
{f_{\Ysssrcsigdisctimecentvecrand}}
%
\newcommand{\ysssrcsigdisctimecentvecranddensval}
{$\Ysssrcsigdisctimecentvecranddensval$}
\newcommand{\Ysssrcsigdisctimecentvecranddensval}
{\Ysssrcsigdisctimecentvecranddens ( \Ysssrcsigdisctimecentvec )}
%
\newcommand{\ysssrcsigdisctimecentindexranddens}
{$\Ysssrcsigdisctimecentindexranddens$}
\newcommand{\Ysssrcsigdisctimecentindexranddens}
{f_{\Ysssrcsigdisctimecentindexrand}}
%
\newcommand{\ysssrcsigdisctimecentindexranddensval}
{$\Ysssrcsigdisctimecentindexranddensval$}
\newcommand{\Ysssrcsigdisctimecentindexranddensval}
{\Ysssrcsigdisctimecentindexranddens ( \Ysssrcsigdisctimecentindex )}
%
%
%
\newcommand{\yssmixsigdisctimecentvecrand}
{$\Yssmixsigdisctimecentvecrand$}
\newcommand{\Yssmixsigdisctimecentvecrand}
{X}
%
\newcommand{\yssmixsigdisctimecentvecranddens}
{$\Yssmixsigdisctimecentvecranddens$}
\newcommand{\Yssmixsigdisctimecentvecranddens}
{f_{\Yssmixsigdisctimecentvecrand}}
%
\newcommand{\yssmixsigdisctimecentvecranddensval}
{$\Yssmixsigdisctimecentvecranddensval$}
\newcommand{\Yssmixsigdisctimecentvecranddensval}
{\Yssmixsigdisctimecentvecranddens ( \Yssmixsigdisctimecentvec )}
%
%
%
%
\newcommand{\ymlsampnb}
{$\Ymlsampnb$}
\newcommand{\Ymlsampnb}
{M}
%
%
%
\newcommand{\ymlcost}
{$\Ymlcost$}
\newcommand{\Ymlcost}
{L}
%
%
%
\newcommand{\ymlcostlnmean}
{$\Ymlcostlnmean$}
\newcommand{\Ymlcostlnmean}
{{\cal L}}
%
\newcommand{\ymlcostlnmeanargmixparam}
{$\Ymlcostlnmeanargmixparam$}
\newcommand{\Ymlcostlnmeanargmixparam}
{\Ymlcostlnmean ( \Ytwoqubitresultphaseevolsin ,
                  \Ysssrcsigdisctimecentone (\Ytwoqubitresultphaseevolsin),
                  \Ysssrcsigdisctimecenttwo (\Ytwoqubitresultphaseevolsin),
                  \Ysssrcsigdisctimecentthree (\Ytwoqubitresultphaseevolsin)
                )}

%
%
%
%
%
%
\newcommand{\ysqrtminusone}
{$\Ysqrtminusone$}
\newcommand{\Ysqrtminusone}
{i}
%
%
%

%% file: artitd47_rapport_v4_no_comment.bbl
\begin{thebibliography}{99}
\bibitem{amoi5-43}
Y. Deville, A. Deville,
"Maximum likelihood blind separation of two quantum states (qubits)
with cylindrical-symmetry Heisenberg spin coupling",
Proceedings of the 2008 IEEE International Conference on
Acoustics, Speech, and Signal Processing (ICASSP 2008),
pp. 3497-3500,
Las Vegas, Nevada, USA,
March 30 - April 4, 2008.
\bibitem{icabook-oja}
A. Hyv\"arinen, J. Karhunen, E. Oja,
"Independent Component Analysis",
Wiley, New York, 2001.
\end{thebibliography}
